\documentclass[reprint,superscriptaddress,preprintnumbers,amsmath,amssymb,aps,prd,nofootinbib]{revtex4-1}

\usepackage{placeins} % Float Barrier
\usepackage[normalem]{ulem} % strikeout
\usepackage{graphicx}  % Include figure files
\usepackage{xcolor}
\usepackage{dcolumn}   % Align table columns on decimal point
\usepackage{bm}        % bold math
\usepackage{hyperref}  % add hypertext capabilities
\usepackage{multirow}
\usepackage{siunitx}
\graphicspath{
  {figures/}
}

\def\ga{\gamma}
\def\de{\delta}

\def\ze{\zeta}

\def\la{\lambda}

\def\si{\sigma}
\def\ta{\tau}

\def\om{\omega}

\def\half{\frac{1}{2}}

\def\bv{{\mathbf{v}}}
\def\bx{{\mathbf{x}}}

\def\bX{{\mathbf{X}}}

\newcommand{\ben}{\begin{equation}}
\newcommand{\een}{\end{equation}}
\newcommand{\bea}{\begin{eqnarray}}
\newcommand{\eea}{\end{eqnarray}}
\newcommand{\ba}{\begin{array}}
\newcommand{\ea}{\end{array}}
\newcommand{\bit}{\begin{itemize}}
\newcommand{\eit}{\end{itemize}}

\newcommand{\IniCorLen}{l_{\phi}}

\newcommand{\ie}{\textit{i.e.\ }}

\newcommand{\Lwind}{\ell_\text{w}}

\newcommand{\ms}{m_\text{sca}} % Scalar mass

\newcommand{\SlenWin}{\ell_\text{w}} % String "winding"length
\newcommand{\SlenRes}{\ell_\text{r}} % String rest frame length
\newcommand{\tStarte}{$\tau_{\rm start}\eta$}
\newcommand{\tEnde}{\ensuremath{\tau_{\rm end}\eta}}
\newcommand{\tDiffe}{$\tau_{\rm diff}\eta$}

\newcommand{\tFit}{\ensuremath{t_{\rm fit}}}
\newcommand{\tcge}{$\tau_{\rm cg}\eta$}
\newcommand{\tStart}{$\tau_{\rm start}$}
\newcommand{\tEnd}{\ensuremath{\tau_{\rm end}}}
\newcommand{\tDiff}{$\tau_{\rm diff}$}
\newcommand{\tcg}{$\tau_{\rm cg}$}
\newcommand{\tOff}{t_0}
\newcommand{\Tc}{T_\text{c}}
\newcommand{\vi}{v}
\newcommand{\vol}{\mathcal{V}}
\newcommand{\vAv}{v}

\newcommand\vev[1]{\left\langle #1 \right\rangle}

\newcommand{\WtFun}{\mathcal{W}}
\newcommand{\ws}{w_0}
\newcommand{\wss}{w}

\newcommand{\xir}{\xi_{\rm r}}
\newcommand{\xr}{x_{\rm r}}

\newcommand{\xiw}{\xi_{\rm w}}

\newcommand{\zzMean}{1.19}
\newcommand{\zzErr}{0.20}
\newcommand{\zzMeansZeroNew}{1.25}
\newcommand{\zzErrsZeroNew}{0.04}
\newcommand{\zzMeansOneNew}{1.20}
\newcommand{\zzErrsOneNew}{0.09}

\begin{document}

\preprint{HIP-2021-7/TH}

\title{Approach to scaling in axion string networks}

\newcommand{\Sussex}{\affiliation{
Department of Physics and Astronomy,
University of Sussex, Falmer, Brighton BN1 9QH,
U.K.}}

\newcommand{\HIPetc}{\affiliation{
Department of Physics and Helsinki Institute of Physics,
PL 64, 
FI-00014 University of Helsinki,
Finland
}}

\newcommand{\EHU}{\affiliation{
Department of Physics,
University of the Basque Country UPV/EHU, 
48080 Bilbao,
Spain
}}

\newcommand{\Tufts}{\affiliation{
Institute of Cosmology, Department of Physics and Astronomy, 
Tufts University,
Medford, MA 02155,
USA}}

\author{Mark Hindmarsh}
\email{mark.hindmarsh@helsinki.fi}
\HIPetc
\Sussex

\author{Joanes Lizarraga}
\email{joanes.lizarraga@ehu.eus}
\EHU

\author{Asier Lopez-Eiguren}
\email{asier.lopez\_eiguren@tufts.edu}
\HIPetc
\Tufts

\author{Jon Urrestilla}
\email{jon.urrestilla@ehu.eus}
\EHU

\date{\today}

\begin{abstract}
We study the approach to scaling in axion string networks in the radiation era, through measuring 
the root-mean-square velocity $v$ as well as the scaled mean string separation $x$.   We find 
good evidence for a fixed point
in the phase-space analysis in the variables  $(x,v)$, providing a strong indication 
that standard scaling is taking place. 
We show that the approach to scaling can be well described by a 
two parameter velocity-one-scale (VOS) model, and show that the values of the parameters 
are insensitive to the initial state of the network. 
The string length has also been commonly expressed in terms of a dimensionless string length density $\zeta$, 
proportional to the number of Hubble lengths of string per Hubble volume. 
In simulations with initial conditions far from the fixed point $\zeta$ is still evolving after half a light-crossing time, which has been interpreted in the literature as a long-term logarithmic growth.   
We show that all our simulations, even those starting far from the fixed point, 
are accounted for by a VOS model with an asymptote of 
$\zeta_*=\zzMeansOneNew\pm\zzErrsOneNew$ (calculated from the string length in the cosmic rest frame) 
and $v_* = 0.609\pm 0.014$.
\end{abstract}

\maketitle

\section{Introduction}
The axion \cite{Weinberg:1977ma,Wilczek:1977pj} is a hypothetical pseudoscalar particle predicted in the Peccei-Quinn (PQ) mechanism \cite{Peccei:1977hh,Peccei:1977ur}. The PQ mechanism extends the Standard Model (SM) of particle physics to solve the so-called strong-CP problem of Quantum Chromodynamics by adding an extra spontaneously broken global U(1) symmetry, which is anomalous.  If the symmetry is broken at a 
high scale  \cite{Kim:1979if,Shifman:1979if,Zhitnitsky:1980tq,Dine:1981rt}, the particle is very weakly interacting and long-lived, and 
becomes a well-motivated dark matter candidate  \cite{Preskill:1982cy,Abbott:1982af,Dine:1982ah}.

The PQ symmetry can be spontaneously broken before or after inflation, leading to very different axionic dark matter scenarios. If the symmetry is broken before inflation, the Universe is filled by a homogeneous axion field, which produces zero-momentum axions through the vacuum realignment mechanism \cite{Preskill:1982cy,Abbott:1982af,Dine:1982ah}. 
On the other hand, in the post-inflationary PQ symmetry breaking, the phase transition happens in the standard cosmology, 
producing axionic cosmic strings \cite{Vilenkin:1982ks,Davis:1986xc}. 
These defects are a variety of global cosmic string \cite{Hindmarsh:1994re,Vilenkin:2000jqa}, which live until the QCD confinement transition, when they form hybrid string-wall composites and are annihilated \cite{Vilenkin:1982ks,Sikivie:1982qv,Georgi:1982ph}. 
The strings formed in this last scenario are the ones studied in this paper. 

Axion strings release energy mainly into pseudo-Goldstone radiation, both during their evolution as well as in the string-wall system collapse.
This radiation constitutes an initially degenerate gas of axions with a non-thermal distribution. 
The complicated non-linear dynamics at the QCD transition imprints density fluctuations which provide 
the seed for the axion minicluster formation through early gravitational collapse \cite{Hogan:1988mp,Kolb:1993zz,Kolb:1994fi,Kolb:1995bu}. 
These miniclusters could be detected by their distinctive small-scale lensing signals \cite{Hogan:1988mp,Kolb:1995bu,Fairbairn:2017dmf,Fairbairn:2017sil}.

The evolution of axion strings and axion radiation is governed by the classical field equations 
of the underlying scalar field theory, and due to their non-linearities, lattice simulations are required 
to go beyond order-of-magnitude estimates. 
In recent years, several groups have studied the evolution of axion strings and the production of axions using lattice simulations 
\cite{Yamaguchi:1998gx,Yamaguchi:1999yp,Yamaguchi:1999dy,Yamaguchi:2002sh,
Hiramatsu:2010yu,Hiramatsu:2012gg,Kawasaki:2014sqa,Fleury:2015aca,Lopez-Eiguren:2017dmc,Klaer:2017qhr,Klaer:2017ond,Gorghetto:2018myk,Kawasaki:2018bzv,Vaquero:2018tib,Buschmann:2019icd,Klaer:2019fxc,Hindmarsh:2019csc,Gorghetto:2020qws,Gorghetto:2021fsn}.   

An accurate calculation of the total number density of axions produced 
is essential for an accurate calculation of the axion energy density, which if matched to the dark matter density today, gives 
a prediction for the axion mass. 
While the axion number density is not very sensitive to the string density at the
QCD transition (see \cite{Dine:2020pds} for a discussion), 
high accuracy in the mass estimate is required for resonant cavity searches, 
which are currently targeting axion masses appropriate for 
production by vacuum realignment \cite{Braine:2019fqb}.

The prediction of the axion density depends on having an accurate description of axion string evolution. 
As the string evolution takes place from the PQ transition at around  $10^{10}$ GeV, to the 
QCD transition at 100 MeV (a factor $10^{11}$), 
the results from numerical simulations  must be 
extrapolated.

It is important to have a physical basis for the extrapolation. Such a physical model is the one-scale model \cite{Kibble:1984hp} and its velocity-dependent improvement 
\cite{Martins:1996jp,Martins:2000cs,Martins:2018dqg}, which we discuss in more detail below.
It predicts that the string network should approach a scaling solution, where the 
mean string separation grows in proportion to cosmic time $t$, and the RMS velocity of the strings is constant.

By scaling we mean that at distances much larger than the string width, 
network length scales such as the mean string separation
are proportional to the cosmic time $t$.  In the  standard 
scaling picture the dynamical evolution of the network 
is independent of the string width and tension. The justification is based on approximating 
the string dynamics by the Nambu-Goto   equations of motion, 
in which the string tension drops out and the string width 
plays no role.

The general picture of network evolution is that strings are initially in a dense tangle of loops and infinite strings, 
with mean separation set by the correlation length of the field and the cooling rate 
\cite{Kibble:1976sj,Zurek:1996sj}. 
They decay by the collapse of loops of string initially present, and chopped off from the infinite strings. 
The mean separation grows,  
until it becomes of order $t$, the cosmic time.  This is approximately as fast as 
causality allows.  By this time the temperature has dropped 
many orders of magnitude, and friction with the cosmic plasma is negligible.  
Strings evolve essentially in vacuum, in the background provided by the the rest of the matter in the universe.

In a previous paper \cite{Hindmarsh:2019csc}, we presented results on 
the scaling dynamics of axion strings. We showed that the network evolution was 
consistent with standard scaling, and 
obtained an asymptotic value for the dimensionless length density of axion string $\zeta$,
which is proportional to the mean number of Hubble lengths of string per Hubble volume, 
$\zeta_\infty = \zzMean \pm \zzErr$. 
This is consistent with, and improves in accuracy, estimates in earlier works 
\cite{Yamaguchi:1998gx,Yamaguchi:1999yp,Yamaguchi:1999dy,Yamaguchi:2002sh,
Hiramatsu:2010yu,Hiramatsu:2012gg,Kawasaki:2014sqa,Lopez-Eiguren:2017dmc}.
Equivalently, the mean string separation $\xi$ is always about half a horizon length.

In \cite{Hindmarsh:2019csc} we also showed how the presence of the scaling solution can be disguised, either by the choice of variable to study, or by the choice of initial conditions.
In this light, claims of a slow or logarithmic growth in the dimensionless length density \cite{Gorghetto:2018myk,Kawasaki:2018bzv,Vaquero:2018tib,Buschmann:2019icd,Martins:2018dqg,Klaer:2019fxc,Gorghetto:2020qws} are to be interpreted as an approach to scaling 
from initially low values of $\zeta$ 
that it is not completed before the simulation ends.

It is important to note that the only significant difference between the simulations of the different groups 
is the method for preparing the initial conditions of the field, 
which determines the initial string separation, 
and that there are no significant differences in the subsequent evolution of the string network. 
All but one group report $\zeta \lesssim 1$ at the end of the simulation, consistent with our estimate. 
The exception \cite{Buschmann:2019icd} explicitly discounts the reliability of their high value, due 
to a non-standard string-finding algorithm.
 
The initial configurations in this work start with random fields with several different initial correlations lengths $\IniCorLen$ in order to cover a range of initial string separations, which 
 tests the sensitivity of the system to the initial conditions. 
The evolution of the system has been carried out using both the true physical field equations, and also using the Press-Ryden-Spergel (PRS) method \cite{Press:1989yh} to simulate strings with constant comoving width.
Simulating (a priori unphysical) strings with constant comoving width allows for a longer period of scaling, thus giving insight into the long-term behaviour of a system of strings. 

We extend the study in  \cite{Hindmarsh:2019csc} by analysing 
the root-mean-square velocity $v$ of the networks alongside 
the mean string separation in units of cosmic time, $x = \xi/t$. 
We demonstrate that the evolution of the simulations at later stages of the simulation 
can be well described by a two-parameter velocity-dependent one-scale (VOS) model 
\cite{Martins:1996jp,Martins:2000cs,Martins:2018dqg} 
where all simulations tend asymptotically to a common point in the phase space $(x_*,v_*)$, 
a fixed point of the VOS dynamical system. 
The dynamical systems analysis predicts that the approach to the fixed point is governed 
by a pair of complex exponents with negative real parts, a stable spiral. 
We find good quantitative accord with the prediction near the fixed point, 
where the model is supposed to be a good description.

Due to this good accord, we obtain a more precise estimate of the scaling density of strings than 
in our previous analysis \cite{Hindmarsh:2019csc}. 
We find that the physical and constant comoving width systems 
have fixed points which are consistent with each other.

Further away from the fixed point, the qualitative agreement is good.  The VOS model predicts that 
initially overdense ($x<x_*$) networks will accelerate, and evolve towards an underdense ($x>x_*$) network as the 
energy-loss mechanism (production and decay of string loops) overcompensates. The approach 
to scaling from the underdense side is a common feature of simulations.
The model also predicts 
that very underdense networks take a long time to reach scaling, often longer than half-box crossing time, consistent with the very underdense simulations of Refs.~\cite{Gorghetto:2018myk,Gorghetto:2020qws}.

%%%%%%%%%%%%%%%%%%%%%%%%%%%%%%%%%%%%%%%%%%%%%%

\section{Model and network parameters}
\label{sec:ModelSims}

\subsection{Field dynamics} 
The simplest axion models include  a singlet scalar field with a U(1) symmetry, $\Phi$, with action
\ben
S=\int d^4 x \sqrt{-g} \Big( \frac{1}{2} \partial_{\mu} \Phi \partial^{\mu}\Phi- \frac{1}{4}\lambda(\Phi^2-\eta^2)^2 \Big),
\label{eq:ac}
\een
where we have written the field as a two-component vector, and the U(1) symmetry is realised as a rotation on the vector.

In a FLRW metric, and when the field is coupled to a thermal bath of weakly-coupled particles, 
the equations of motion take the form
\ben
{\Phi}''+2\frac{{a'}}{a}{\Phi'}-\nabla^2 \Phi = -a^2 \lambda (\Phi^2-\eta^2(T))\Phi,
\label{eq:eom}
\een
where $a$ is the scale factor, a prime denotes differentiation with respect to conformal time $\tau$, and 
in the radiation era $a \propto \tau$. 
The free energy of the system is minimised at the field magnitude $\eta(T)$, where 
$\eta^2(T) = d(\Tc^2 - T^2)$, 
$\Tc \simeq \eta$ is the critical temperature of the PQ phase transition, and $d$ is a constant computable in perturbation theory.
For $T \gg \Tc$, it is energetically favourable for the field to fluctuate around $\Phi = 0$. 
Well below the critical temperature it is energetically favourable for the magnitude of the field to take the value $\eta$, 
with a massless pseudoscalar fluctuation mode (the axion) and a scalar mode of mass $\ms = \sqrt{2\la}\eta$.
During the phase transition, the direction in field space is  chosen at random in uncorrelated regions of the universe, 
with the result that the field is forced to stay zero along lines \cite{Kibble:1976sj}.  These lines form 
the cores of the axion strings \cite{Davis:1986xc}. 
The size of the core is approximately $\ws = \ms^{-1}$.

\subsection{Network parameters from field averages}
\label{sec:ScObs}
%

%%%%%%%%%%%%%%%%%%%%%%%%%%%%%%%%%%%%%%%%%%%%%%

The subsequent evolution of the string network can be tracked by the 
string length $\ell$ and the RMS velocity $\vAv$ of the strings. 

A couple of estimators for $\ell$ are possible.
We define the winding length $\Lwind$ 
as the number of plaquettes pierced by strings multiplied by the 
physical lattice spacing $a\delta x$,  
corrected by factor of $2/3$ to compensate for the Manhattan effect \cite{Fleury:2015aca}. 
Such plaquettes are identified calculating the ``winding'' phase of the field around each plaquette 
of the lattice \cite{Vachaspati:1984dz}. 
This is an estimate of the length of string measured in the ``universe frame'', that is, observers 
comoving with the expansion of the universe. 

Other measures of length are based on the observation that the energy of a string configuration is proportional to its 
length, and the estimators are constructed using local functions of the fields.  
To simplify the discussion, we will first neglect the expansion of the universe.

Consider a weighted total energy
\ben
E = E_\pi + E_{D} + E_V
\een
with the functions 
\bea
E_{\pi} &=& \half \int d^3 x \Pi^2 \WtFun(\Phi), \\
E_{D} &=&  \half \int d^3 x (\nabla\Phi)^2 \WtFun(\Phi), \\
E_V &=& \int d^3 x V(\Phi) \WtFun(\Phi),
 \eea
where $\Pi = (\partial_t \Phi)$  and $V(\Phi)=\frac{1}{4}\lambda(\Phi^2-\eta^2)^2$. 
The function $\WtFun(\Phi)$ is a local function of the fields which is strongly peaked near $\Phi = 0$, 
and zero for $|\Phi| = \eta$, so that it picks out strings.
We call the three functions defined above the weighted kinetic, gradient and potential energy respectively.

Suppose that all the energy in the volume $\vol$ is in the form of global strings, centered on the line $\bX(\si,t)$. 
The coordinate $\si$ is chosen so that 
\(|\bX'| = (1 - \dot\bX^2)^\half ,\) where the prime represents the derivative with respect to $\si$, and the dot the derivative with respect to $t$.
We  denote the total rest-frame length of string 
\ben
\SlenRes = \int d\si.
\een 
Writing local rest frame space coordinates $\bx_\text{s}$, and 
fields measured in the local rest frame with the subscript s,  
the fields of a piece of string moving with orthogonal velocity $\dot \bX$ are  
\bea
\Pi(\bx,t) &=& \ga \dot\bX \cdot \nabla \Phi_\text{s}(\bx_\text{s}), \\
\nabla\Phi(\bx,t) &=&  \ga \hat{\bv} (\hat{\bv} \cdot \nabla \Phi_\text{s}(\bx_\text{s}) ) + \nabla^\perp\Phi(\bx,t),
\eea
where $\hat\bv$ is a unit vector in the direction of $\dot\bX$, $\ga =1/\sqrt{1 - \dot\bX^2}$ is the boost factor,
and 
\ben
\nabla^\perp_i\Phi(\bx,t) = (\de_{ij} - \hat{v}_i\hat{v}_j) \nabla_j \Phi(\bx,t).
\een
Choosing the local rest frame so that the string is 
oriented in the $z_\text{s}$ direction, a string moving with velocity $\dot\bX$ has scalar kinetic energy
\bea
E_{\pi} &=& \frac14 \int dx_\text{s}dy_\text{s} (\nabla\Phi_\text{s})^2 \WtFun(\Phi_\text{s}) \int d\si \dot\bX^2\,,
\label{e:Epi}
\eea
  gradient energy  
\bea
E_{D} &=& \frac14 \int dx_\text{s}dy_\text{s} (\nabla\Phi_\text{s})^2\WtFun(\Phi_\text{s})   \int d\si \left( 1  + \frac{1}{\ga^2} \right),
\label{e:ED}
\eea
and potential energy 
\ben
E_V = \int dx_\text{s}dy_\text{s}V(\Phi_\text{s})\WtFun(\Phi_\text{s})   \int d\si \frac{1}{\ga^2} .
\label{e:EV}
\een
The total energy is therefore 
\ben
E = \mu ( 1 - f_V \vAv^2 ) \SlenRes,
\een
where 
\bea
\mu &=&  \int dx_\text{s}dy_\text{s} \left[ \half (\nabla\Phi_\text{s})^2 \WtFun(\Phi_\text{s}) 
+ V(\Phi_\text{s}) \WtFun(\Phi_\text{s})  \right]
\label{e:StrMuDV}
\eea
is the $\WtFun$-weighted mass per unit length of a static string,
with $f_V$ the fraction contributed by the potential energy density, 
and we have 
defined an RMS velocity $\vAv$ through
\ben
\vAv^2 = \frac{1}{\SlenRes} \int d\si \dot\bX^2 .
\een
A convenient choice for the weight function is 
\ben
\WtFun = V(\Phi) ,
\een
for which $\mu = 0.892\eta^2$ and $f_V = 0.368$.\footnote{These numbers are obtained from a code implementing a relaxation method on the discretised radial energy functional, with $800$ lattice points and lattice spacing $\eta dr = 0.01$. The convergence criterion was that the change in energy in an update should be less than $10^{-5}\eta$. }

Besides the energy, we can also calculate the Lagrangian 
\bea
L &=& E_\pi - E_{D} - E_V, \\
\eea
finding
\bea
L &=& -\mu ( 1 - \vAv^2)\SlenRes. \label{e:ASlag} 
\eea
Combining the total energy $E$ and the Lagrangian estimators, an estimate for both the rest-frame 
length $\SlenRes$ and the mean square 
velocity can be obtained
\bea
\SlenRes &=& \frac{E + f_V L}{\mu(1 - f_V)} \label{eq:slenres}, \\
\vAv_L^2 &=& \frac{E + L}{E + f_VL},
\eea
where the subscript $L$ denotes the use of the Lagrangian to obtain the estimate. 
An alternative way of estimating the string velocity comes from the pressure, 
\ben
p\vol = E_\pi - \frac{1}{3}E_{D} - E_V,
\label{eq:vel_lag}
\een
which depends on the rest frame length and RMS velocity as 
\ben
p\vol=\frac{1}{3}\mu\SlenRes\left[ (2v^ 2-1)-f_V(2-v^2)\right] .
\een
It is then straightforward to derive another mean square velocity estimator
\ben
\vAv_\om^2 = \frac{1 + 3\om + 2f_V}{2 + f_V(1 + 3\om) }, 
\label{eq:vel_w}
\een
where $\om = p\vol/E$ is the equation of state parameter of the strings. 
A third estimate for the string velocity can be constructed from the ratio of the kinetic to gradient energies \cite{Hindmarsh:2017qff}, 
\ben
R_{\text{s}} = \frac{E_{\pi}}{E_{D}},
\een
which can be rearranged to give 
\ben
\vAv_{\text{s}}^2 = \frac{2R_{\text{s}} }{1+ R_{\text{s}} }.
\label{eq:vel_s}
\een
Given that we only have   three independent underlying quantities $E_\pi$, $E_D$, and $E_V$, 
only three independent estimators can be derived from them: one length, and two velocity estimators. 

The winding length is not derived from the weighted energies, and so is an independent length estimator. 
As it is the ordinary Euclidean length of the curve traced by the string, 
it can be represented as 
\ben
\label{e:LenWinRes}
\SlenWin = \int d\si |\bX'| = \SlenRes \left\langle \ga^{-1} \right\rangle .
\een
Note that the average of $\gamma^{-1}$ is not in general equal to $(1 - \vAv^2)^{1/2}$.

In a cosmological simulations one can express the string length in terms of Hubble lengths per Hubble volume, or 
\ben
\label{e:ZetDef}
\zeta = \frac{\ell t^2}{\vol} .
\een
When investigating scaling in string networks, it is more transparent to parametrise the string density 
by the mean string separation, which is obtained from measures of the string length via
\ben
\label{e:XiDef}
\xi = \sqrt{\frac{\vol}{\ell}} .
\een
In this work we will use two length estimators, which will define two different mean string separation estimators: when the length estimator used is the rest-frame estimator  $\SlenRes$, we will define $\xir$; and when the length estimator used is the winding length estimator  $\SlenWin$, we will define $\xiw$.

The above estimators were derived for a Minkowski space-time.  In an expanding background, one can view the space-time 
coordinates as representing comoving position and conformal time, from which physical lengths follow by multiplication 
by the scale factor $a$.

%%%%%%%%%%%%%%%%%%%%%%%%%%%%%%%%%%%%%%%%%%%%%%
\section{Simulations and scaling observable results}

We solve a discretised version of the equations of motion (\ref{eq:eom}) on a cubic lattice with periodic boundary conditions, evolving the system in conformal time $t$
The results we present in this section are extracted from the same set of simulations analysed in \cite{Hindmarsh:2019csc}, where lattices with $4096$ sites per dimension were used with spatial resolution of $\delta x\eta=0.5$ and conformal time steps of $\delta \tau\eta=0.1$. In addition, we use a set of simulations with a larger initial correlation length, but otherwise identical. 
In the following lines we will only summarise the procedure and refer the reader to \cite{Lopez-Eiguren:2017dmc,Hindmarsh:2019csc} for more detailed descriptions on the method.

The field configuration is initiated at conformal time \tStart\ by setting the scalar field canonical momentum to zero and the scalar field to be a Gaussian random field with power spectrum $P_{\Phi}(k)={A}\left[{1+(k\IniCorLen)^2}\right]^{-1}$, were $A$ is chosen so that $\langle \Phi^2 \rangle=\eta^2$ and $\IniCorLen$ is the field correlation length in comoving coordinates.  We use different values of $\IniCorLen$ in order to cover a range of string separations in the initial conditions.
In order to allow the strings to form, and to remove the energy excess in the field fluctuations around the string configurations, we evolve this configuration with a diffusion equation 
with unit diffusion constant until conformal time \tDiff. 
We then apply the second-order time evolution equation (\ref{eq:eom}).

Similarly to our previous paper, we extract data from simulations with both fixed comoving string width and fixed physical string width. 
We promote the scalar self-coupling constant to be a time dependent parameter $\la = \la_0/a^{2(1-s)}$ following the  PRS  method \cite{Press:1989yh}. This makes the comoving string width decrease 
with conformal time as:
\begin{equation}
\wss(\tau) = \frac{\ws}{a^s(\tau)} \, .
\label{e:StrWid}	
\end{equation}
The physical equation of motion, where the physical string width remains constant at $\ws = 1/\sqrt{2\la_0}\eta$, and 
the comoving width decreases with time, corresponds to $s=1$. 
With $s=0$ the comoving width is constant at $\ws$ and the physical 
string width increases in time. 

For the $s=1$ case, 
it is difficult to avoid the string width being larger than the Hubble length at early times, 
which also means that the relaxation of the field to its equilibrium value is longer than a Hubble time. 
In order to speed up the string formation, 
we arrange the time-dependence of the coupling so that strings are formed and diffused 
with a constant comoving width, equal to their final comoving width. 
At the end of the diffusion period, 
the string width is much smaller than its physical value $\ws$. 
The string width is then allowed to grow 
by setting $s = -1 $ until \tcg, 
which is when the string core has expanded to its correct physical width $\ws$. 
After conformal time \tcg, the physical evolution with $s=1$ starts. 
We call this procedure core growth. 
Simulations end at conformal time \tEnd.

Table \ref{tab:sims} contains all simulation parameter choices that have been considered in the procedures described above.  
Four simulations with different random number seeds were carried out at each parameter choice, for a total of 28 runs. 
The data are analysed in cosmic time $t = (\tau/\tEnd)^2\tEnd/2$.

\begin{table}[h!]
\renewcommand{\arraystretch}{1.5}
\begin{tabular}{|c||c||c|}
\hline
 Model & $s=1$ & $s=0$  \\\hline
 $\IniCorLen\eta$ & (5,10,20,40) & (5,10,20) \\ \hline
 \tStarte & 50 & 50  \\
 \tDiffe & 70 & 70 \\
 $s_{\rm cg}$ & -1 & -- \\
 \tcge & 271.11 & -- \\
 \tEnde & 1050 & 1050  \\
 \hline
\end{tabular}
 \caption{\label{tab:sims} Run parameters used in simulations. See text for explanation.}
\end{table}

\begin{figure}[htbp]
    \centering
    \includegraphics[width=\columnwidth]{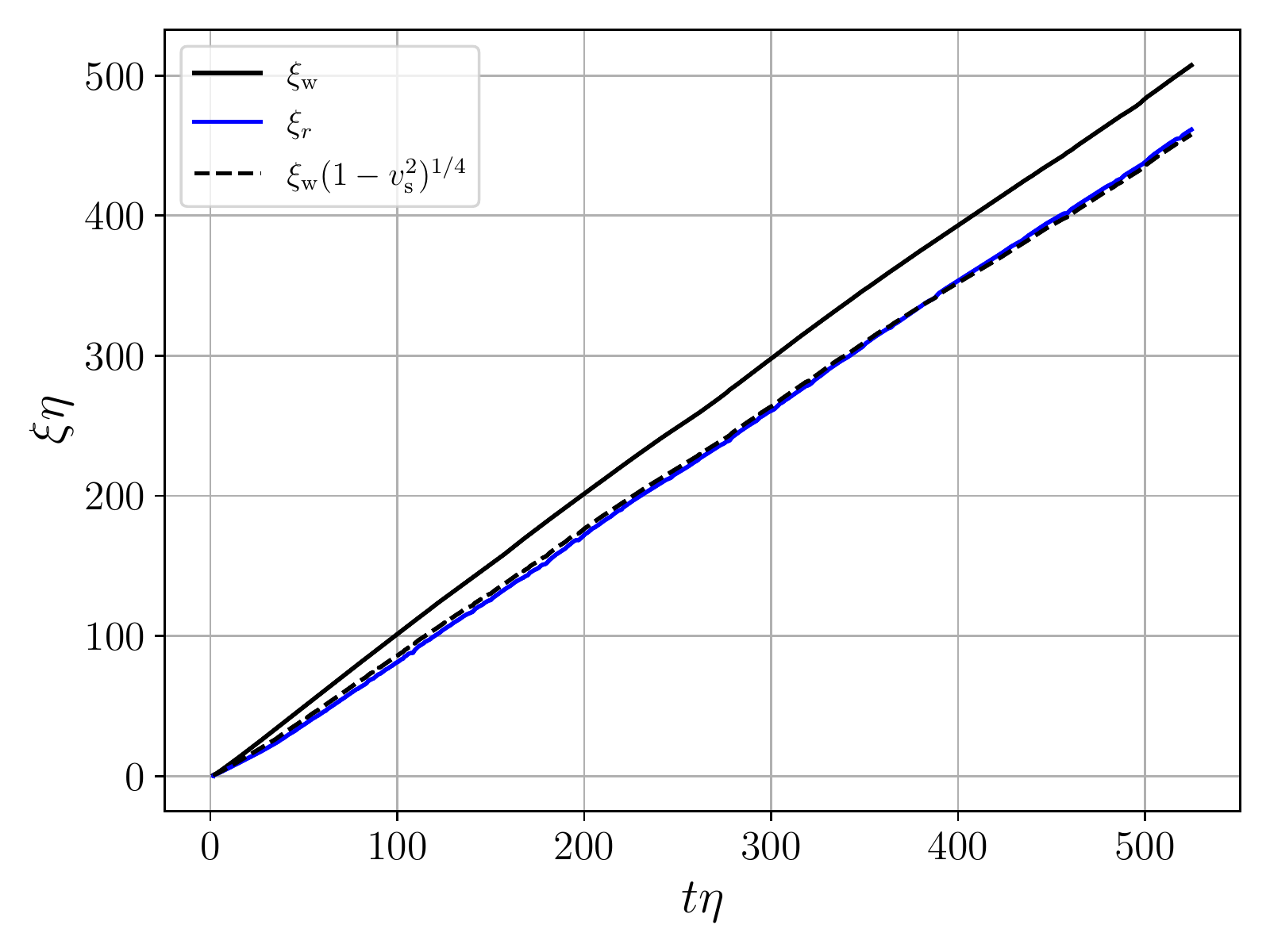}
    \includegraphics[width=\columnwidth]{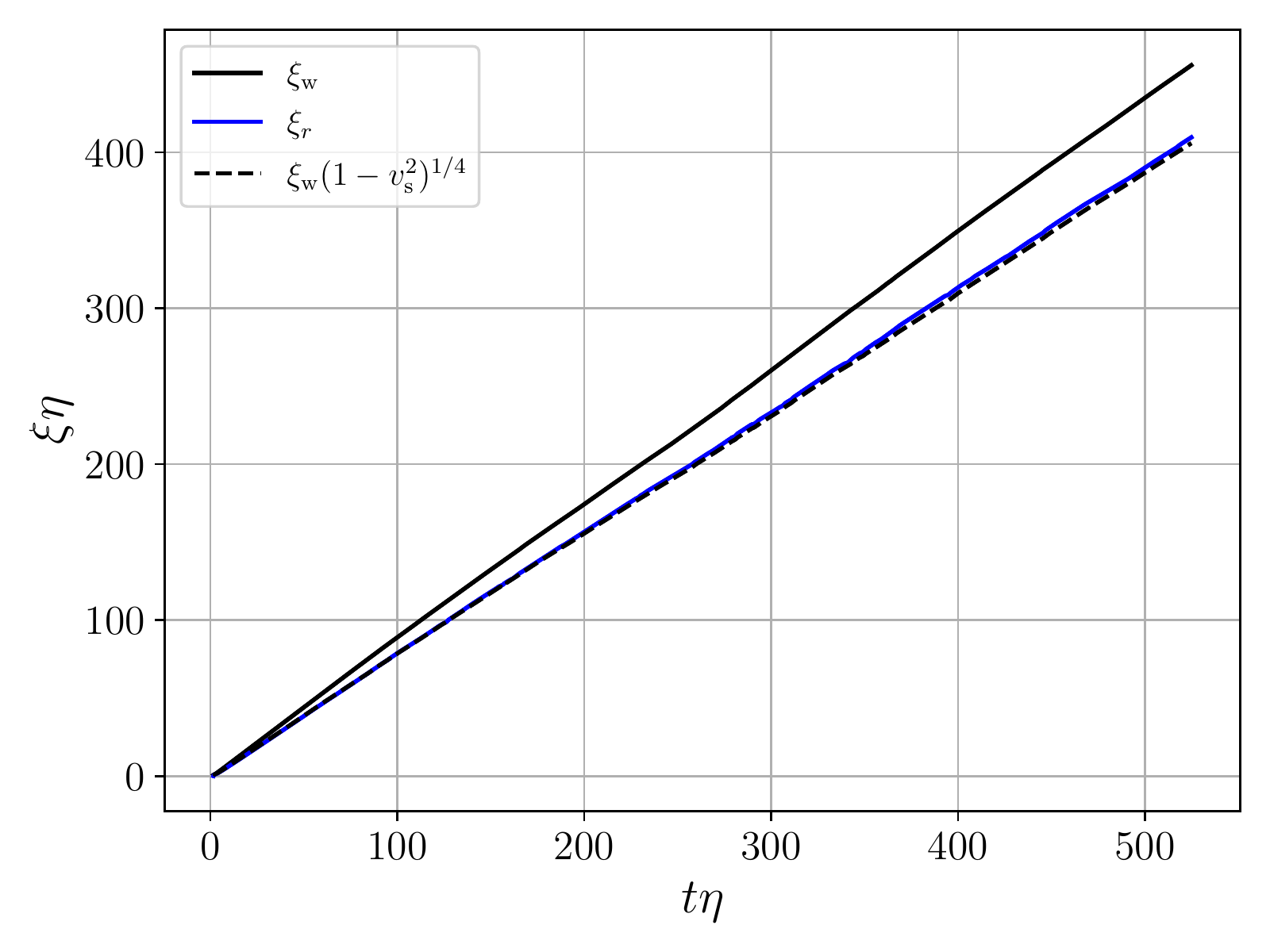}
    \caption{
     Comparison of the mean string separation defined from  the winding length estimator $\xiw$ (solid black) and the rest-frame estimator $\xir$ (solid blue) as presented after Eq.~(\ref{e:XiDef}), from a single simulations with correlation length $l_{\phi}\eta=5$ and $s=1$ (upper panel) and $s=0$ (lower panel).  The dashed black line corresponds to the winding estimator $\xiw$ modified via Eq.~(\ref{e:LenWinRes}).
    \label{fig:xiScaling_comp}}
 \end{figure}
 
 \begin{figure}[htbp]
    \centering
    \includegraphics[width=\columnwidth]{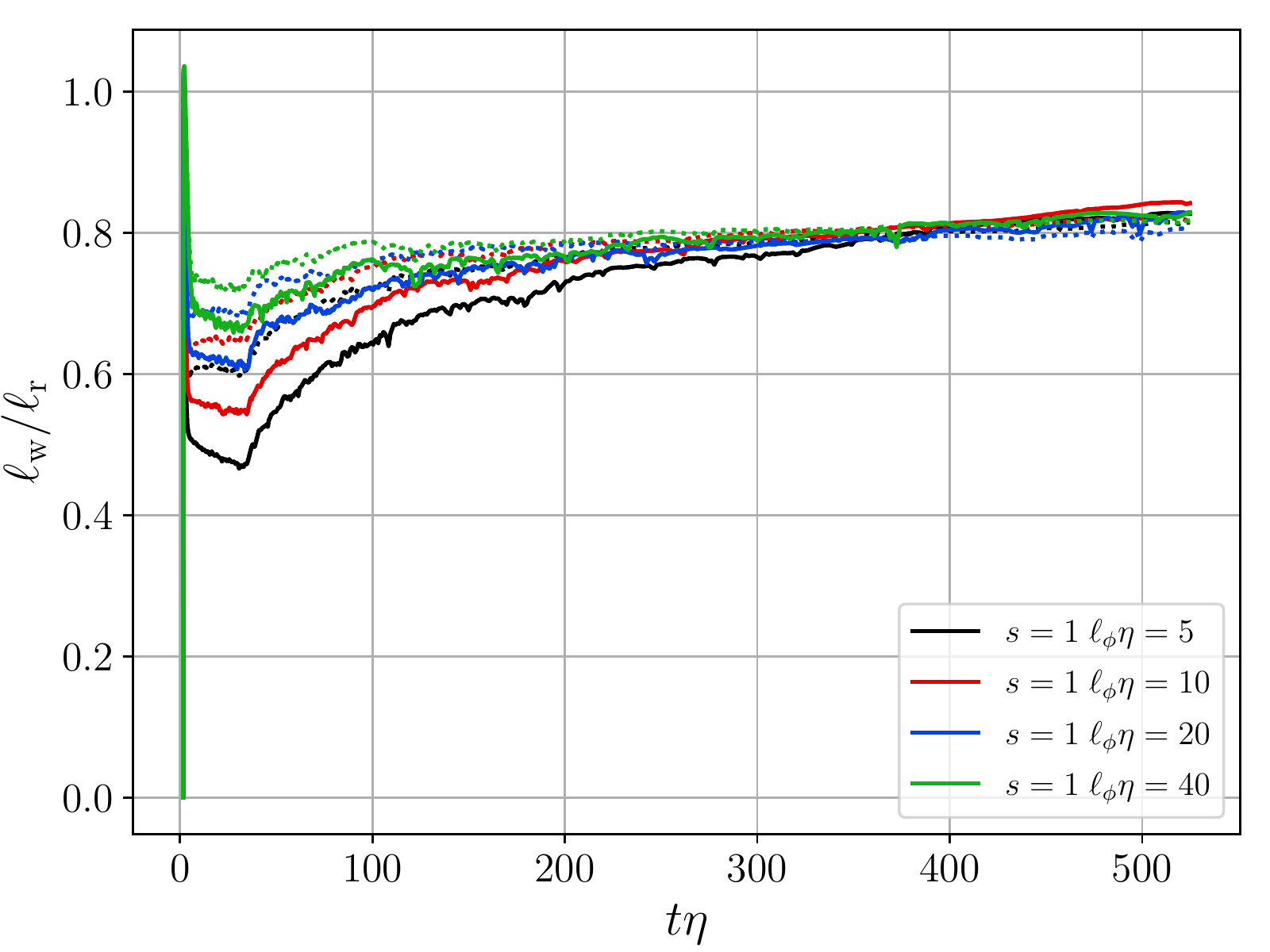}
    \includegraphics[width=\columnwidth]{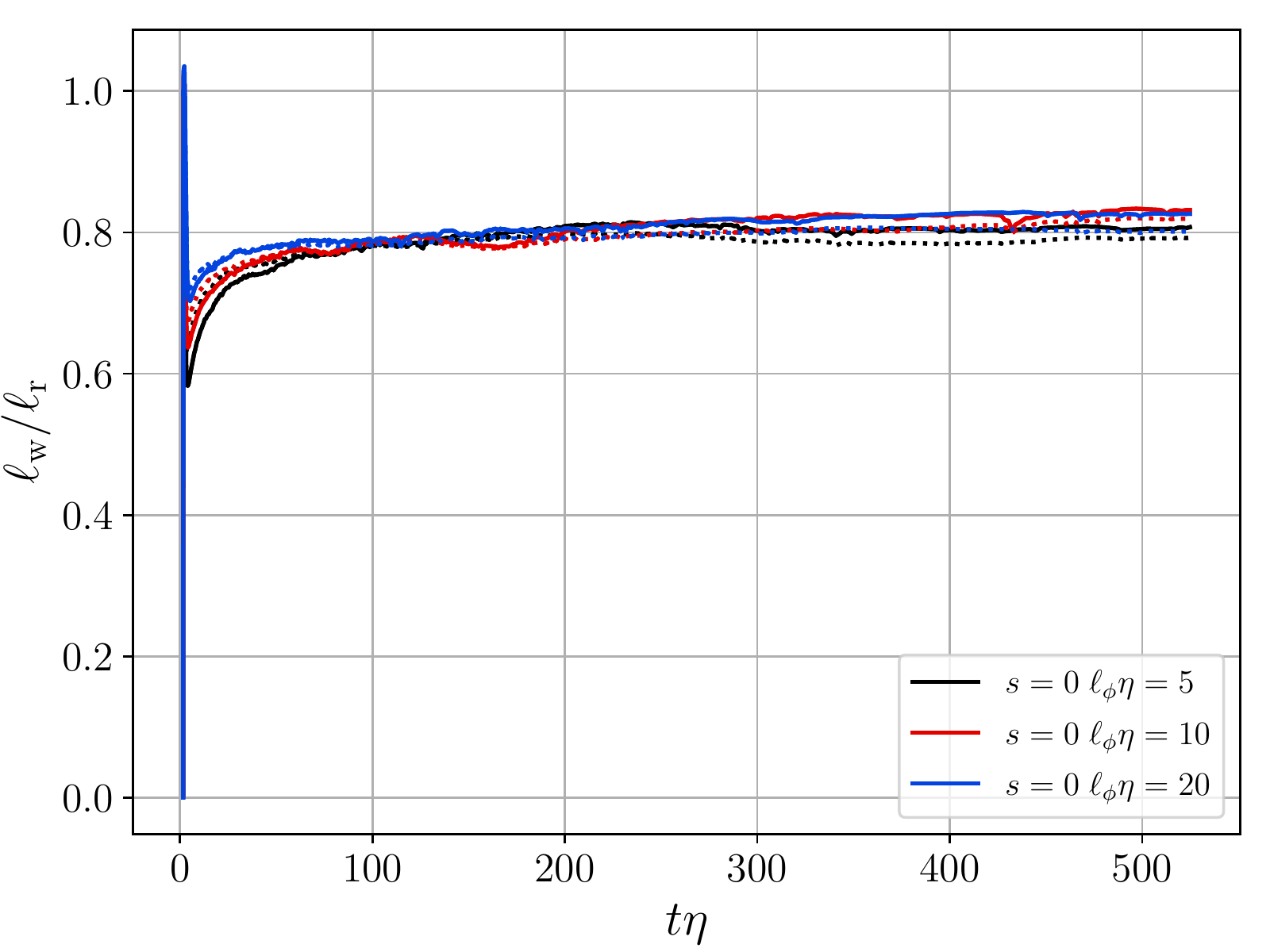}
    \caption{
     Ratios of the winding length estimator $\SlenWin$ to rest-frame estimator $\SlenRes$ (solid line) for different correlation lengths. The lines correspond to a single simulations. In dashed, we plot $(1 - \vAv^2_\text{s})^{1/2}$.
    \label{fig:xiratio}}
 \end{figure}
 
  \begin{figure}[htbp]
    \centering
    \includegraphics[width=\columnwidth]{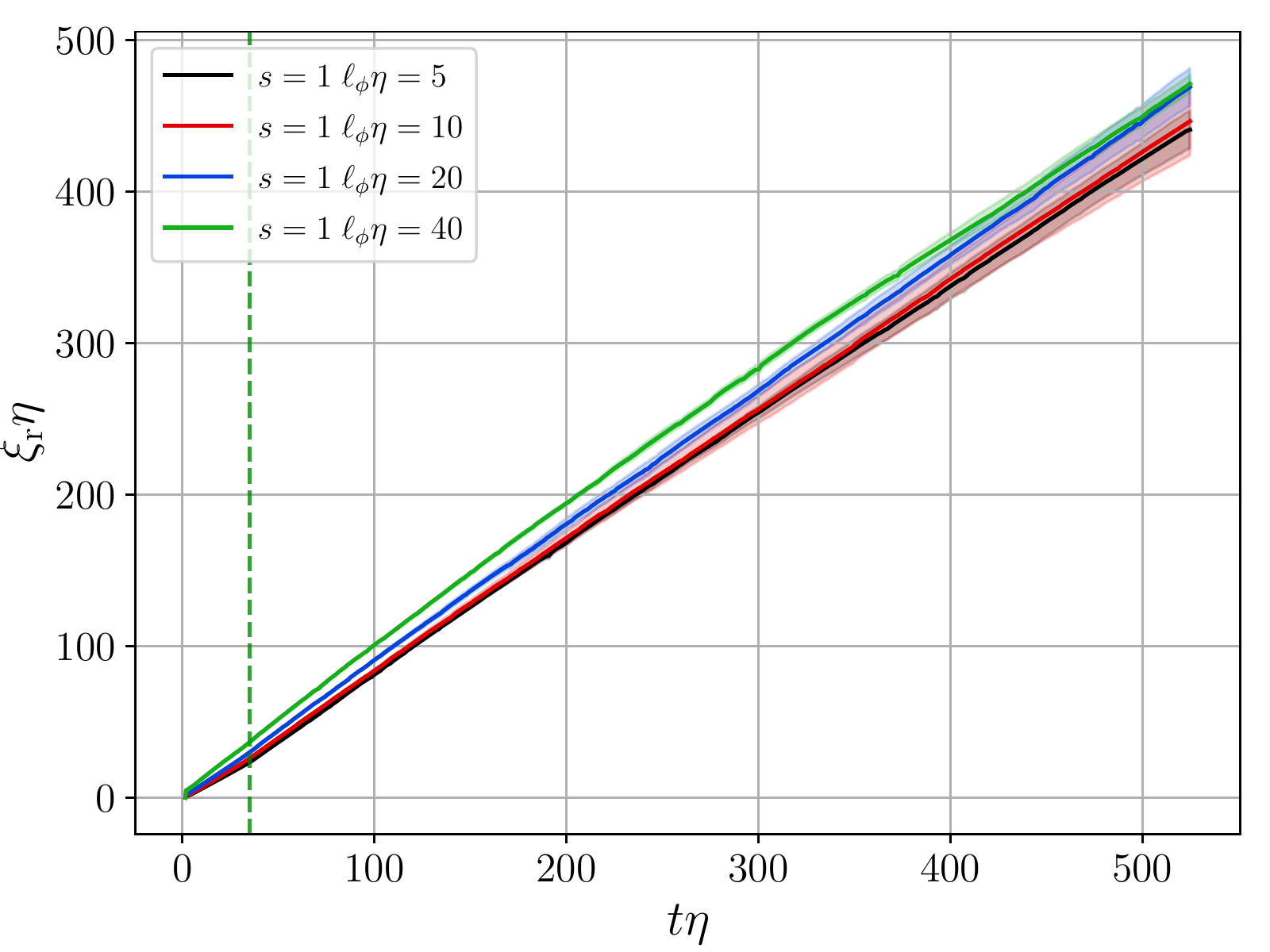}
    \includegraphics[width=\columnwidth]{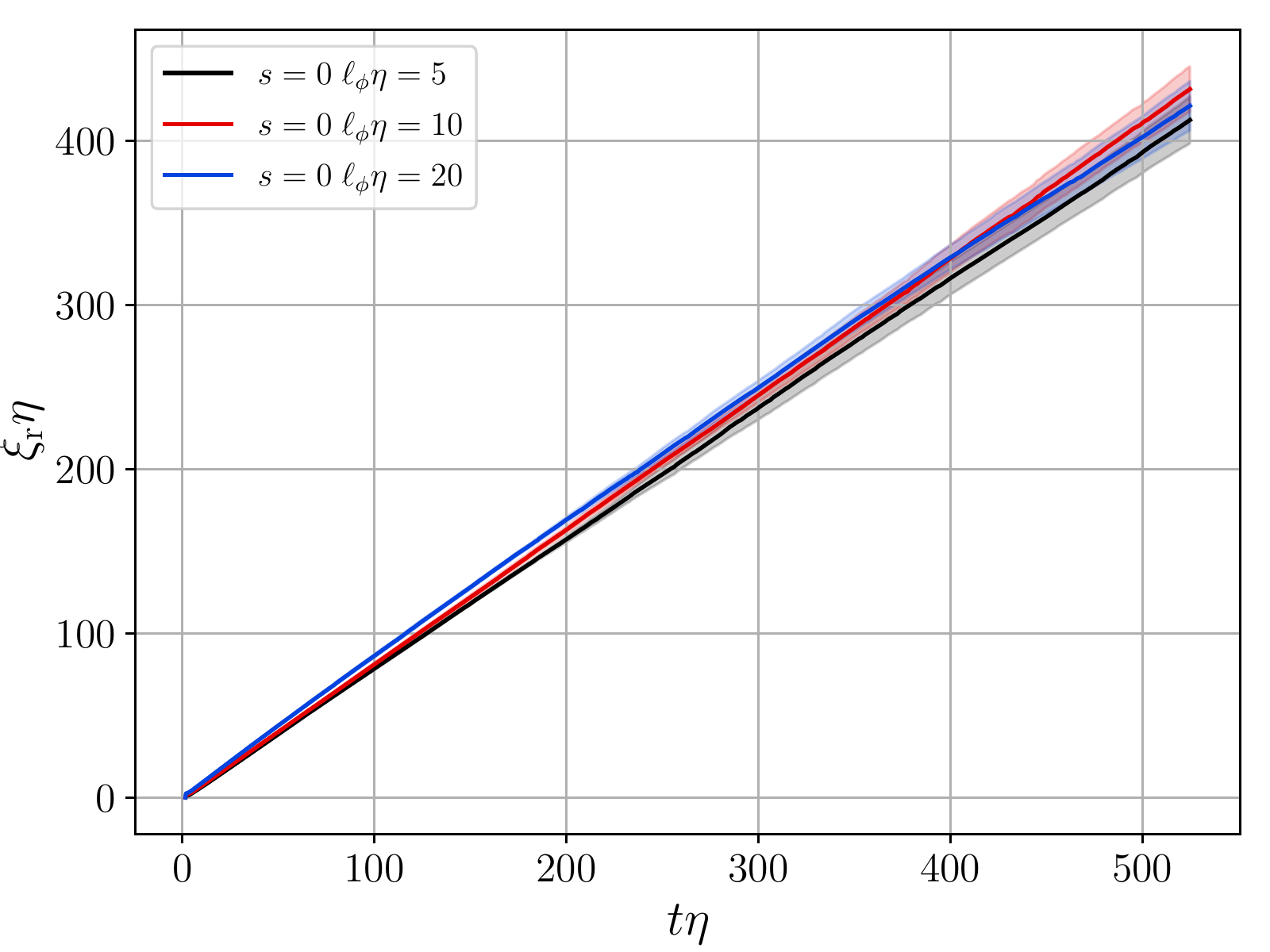}
    \caption{
    Mean string separation $\xi_r$ from simulations with $s=1$ (top panel) and $s=0$ (bottom panel)  with initial field   
    correlation lengths $\IniCorLen \eta=5$ (black), $\IniCorLen \eta=10$ (red), $\IniCorLen \eta=20$ (blue) and $\IniCorLen \eta=40$ (green - only for $s=1$). 
    The solid line represents the mean over realisations of $\xi$ at each time, with 
    the shaded regions showing the 1-$\sigma$ variation. The vertical green line corresponds to the end of the core growth period, 
    after which the system is evolved with the physical equations of motion in the $s=1$ case. 
    \label{fig:xiScaling}}
 \end{figure}

Figure \ref{fig:xiScaling_comp} shows the comparison of the evolution of the mean string separation for $\xir$ and $\xiw$ presented in the previous section (\ref{e:XiDef}) for a single run with $\IniCorLen\eta=5$. 
The upper panel is for simulations with $s=1$ and the lower panel for $s=0$. Their growth is consistent with a linear asymptote, as extensively studied in \cite{Hindmarsh:2019csc}. This is the expectation from the standard picture of scaling in axion string networks 
\cite{Vilenkin:1982ks,Kibble:1984hp,Martins:1996jp}. 
Note that in \cite{Hindmarsh:2019csc}, the winding length estimator $\xiw$ was used to establish the asymptotic linear growth; here we establish that the rest-frame estimator also grows linearly, as expected.

The ratio of the winding length estimator $\SlenWin$ to the rest frame estimator $\SlenRes$ is plotted in Fig.~\ref{fig:xiratio} (solid line), which according to Eq.~(\ref{e:LenWinRes}) 
is an estimate of $\vev{\ga^{-1}}$, where $\ga$ is the Lorentz factor of the string.  
For comparison, we plot $(1 - \vAv^2_\text{s})^{1/2}$, using the scalar field estimator (\ref{eq:vel_s}), whose 
time-dependence in the simulations is discussed later in this section. 
As pointed out in the previous section, the two quantities are not necessarily equal, but empirically we observe that 
they are close by the end of the simulation.

The closeness of $(1 - \vAv^2_\text{s})^{1/2}$ to $\vev{\ga^{-1}}$ is also observed in Fig.~\ref{fig:xiScaling_comp},  where we show as a dashed line the winding estimator multiplied by $(1 - \vAv^2_\text{s})^{1/2}$. 
We choose $\SlenRes$ as the length estimator for the rest of this work, 
which is better suited to the dynamical modelling we carry out.  It can be related to the winding length through the factor of approximately $0.8$ on show in Fig.~\ref{fig:xiratio}.

The evolution of the mean string separation for all simulations is shown in Fig.~\ref{fig:xiScaling}. The solid line represents the mean obtained by averaging over realisations and the shaded regions the $1\sigma$ standard deviations. 
Uncertainties are calculated by propagating the fluctuations in the weighted energies \eqref{e:Epi}, \eqref{e:ED} and \eqref{e:EV}.
We use black for $\IniCorLen\eta=5$,  red for $\IniCorLen\eta=10$, blue for $\IniCorLen\eta=20$, and green for $\IniCorLen\eta=40$ (only in simulations with $s=1$). 
The end of the core growth period  is shown as a vertical green dashed line. 
These figures extend the results of Fig.~1 in Ref.~\cite{Hindmarsh:2019csc}, which shows the winding length estimator only for $s=1$, 
and a subset of the initial correlation lengths.

 \begin{figure}[htbp]
    \centering
    \includegraphics[width=\columnwidth]{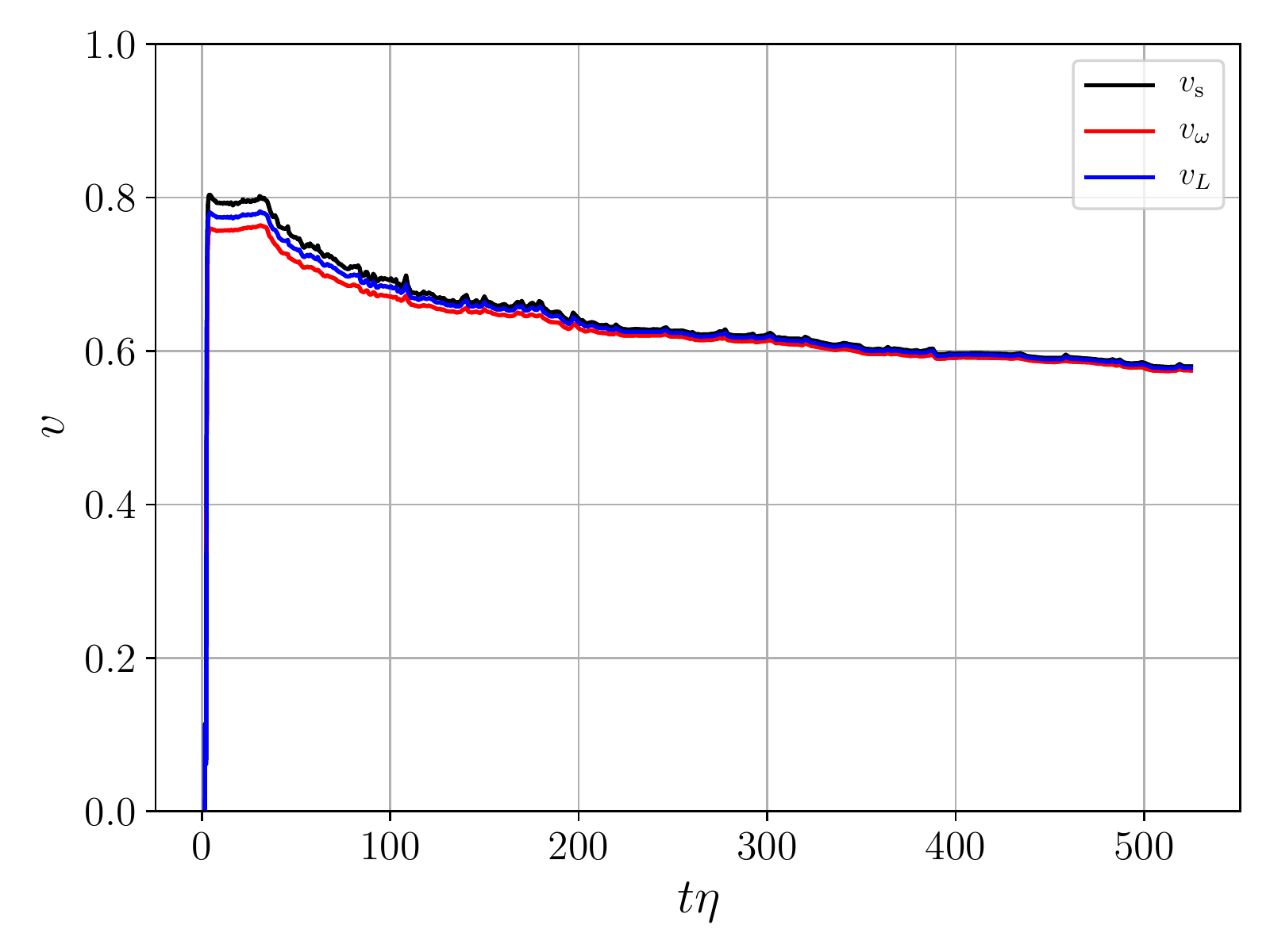}
    \includegraphics[width=\columnwidth]{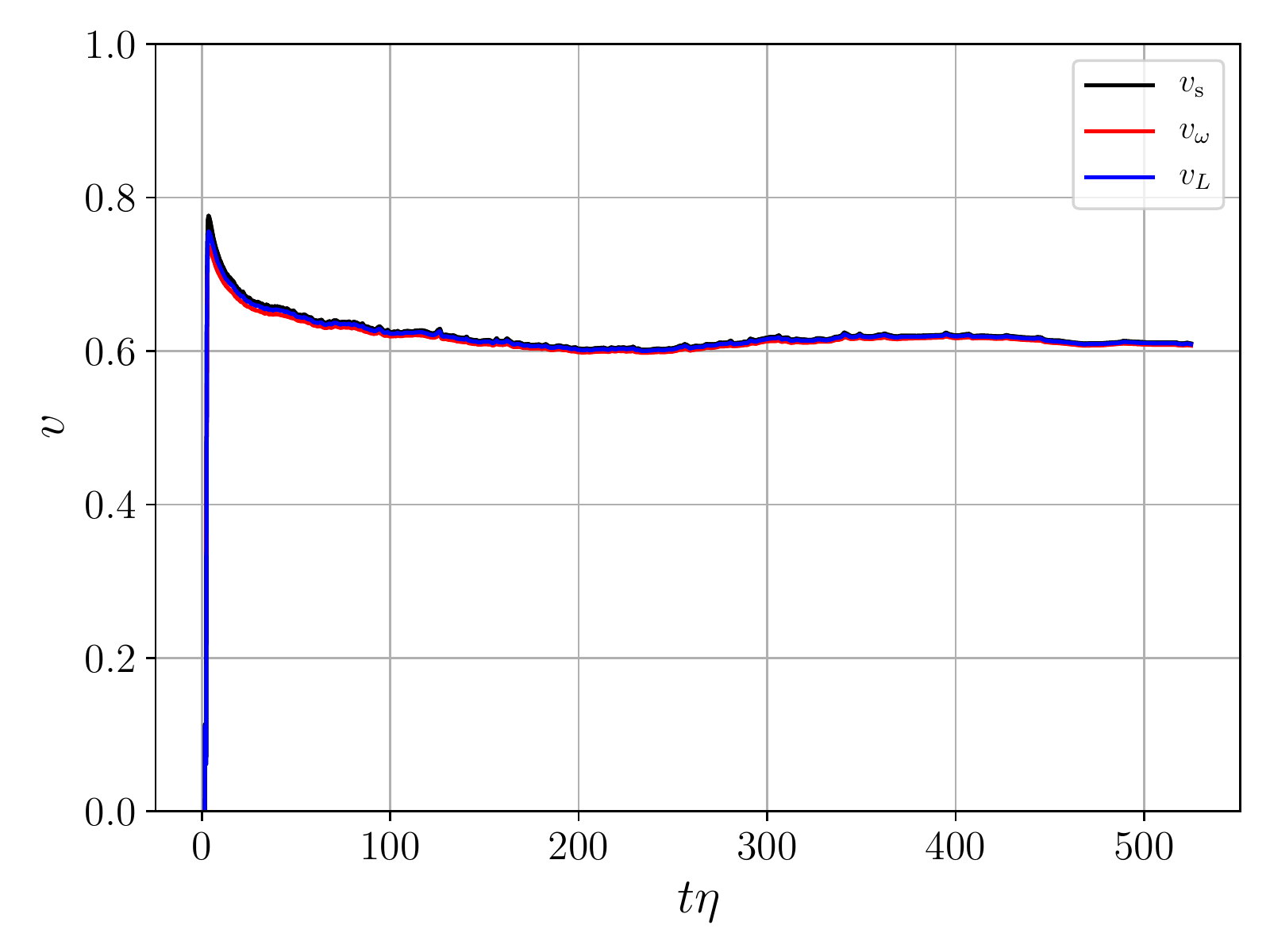}
    \caption{ 
    Comparison of velocity estimators presented in Sec.~\ref{sec:ScObs} for a simulation with correlation length $\IniCorLen \eta=5$ and $s=1$ (upper panel) and $s=0$ (lower panel). The values of the scalar field velocity estimator  $\vAv_{\text{s}}$  (\ref{eq:vel_s}), the equation of state velocity estimator  $\vAv_\om$ (\ref{eq:vel_w})  and the Lagrangian-derived velocity estimator $\vAv_L$ (\ref{eq:vel_lag}) are shown in  black, red and blue, respectively.
    \label{fig:v_comp}}
 \end{figure}
 
  \begin{figure}[t]
    \centering
    \includegraphics[width=\columnwidth]{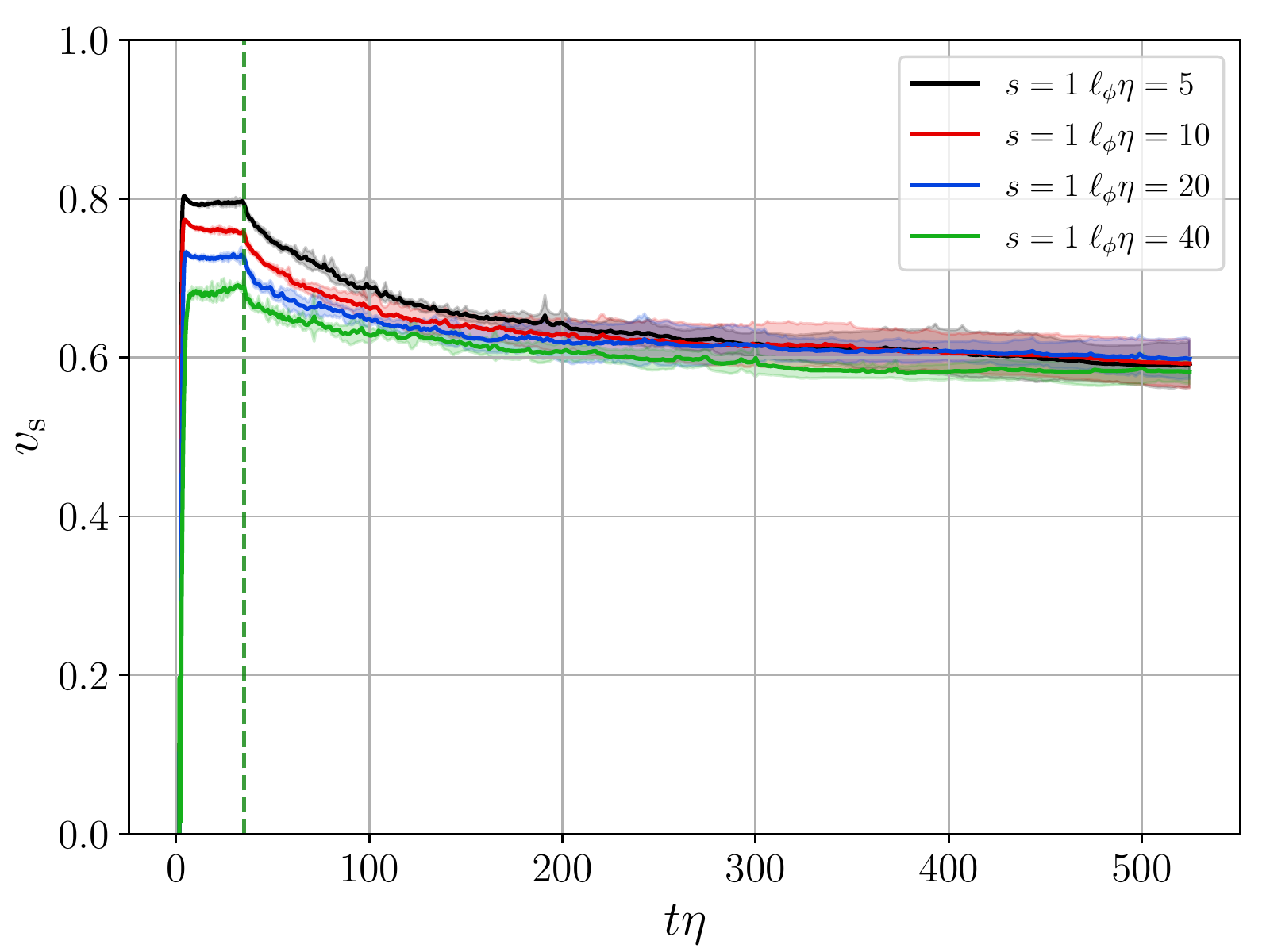}
    \includegraphics[width=\columnwidth]{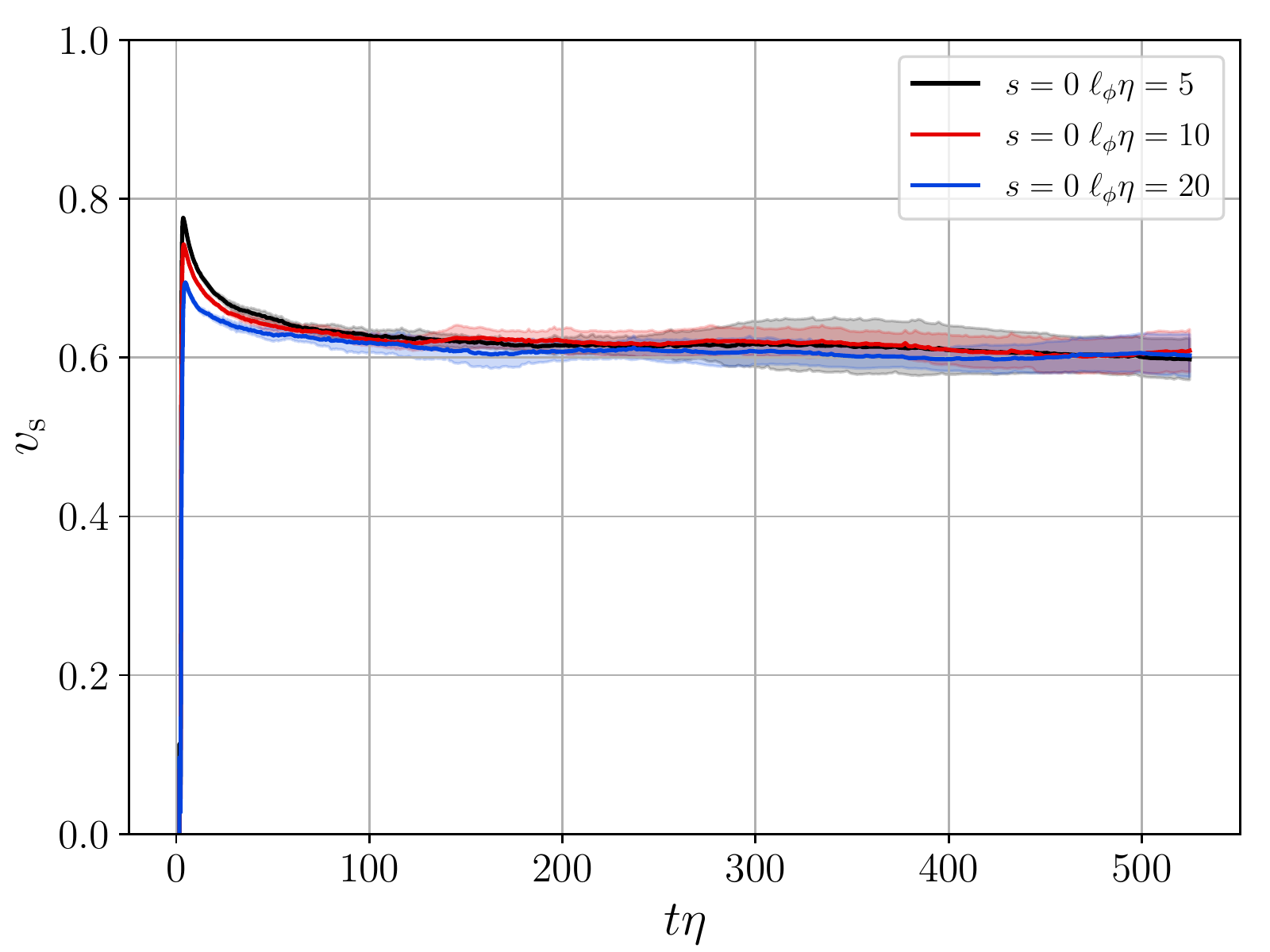}
    \caption{ 
    Velocities from scalar field estimator $v_{s}$ (\ref{eq:vel_s}) from  simulations with $s=1$ (top panel) and $s=0$ (bottom panel) with initial field   correlation lengths $\IniCorLen \eta=5$ (black), $\IniCorLen \eta=10$ (red), $\IniCorLen \eta=20$ (blue) and $\IniCorLen \eta=40$ (green - only for $s=1$). 
    The solid line represents the mean over realisations at each time, with 
    the shaded regions showing the 1-$\sigma$ variation. 
    The vertical green line corresponds to the end of the core growth period. 
    \label{fig:v_s_Scaling}}
 \end{figure}

We now turn to the velocity estimators. 
To establish their consistency, we plot all three for the same run in Fig.~\ref{fig:v_comp}, 
with $s=1$ in the top panel and $s=0$ in the bottom panel.  
As mentioned in the previous section, only two are independent, but the fact that all three 
are so close gives confidence that they are indeed estimating a global translational velocity of a 
string-like solution, rather than field fluctuations in regions where $\Phi$ is close to zero, which is 
a potential contaminant of velocity estimators.

Uncertainties in velocities 
are calculated by propagating the fluctuations in the weighted energies \eqref{e:Epi}, \eqref{e:ED} and \eqref{e:EV}.
We find that the 
largest fluctuations are in the weighted potential energy $E_V$.  We therefore choose the 
estimator $\vAv_\text{s}$ derived from kinetic and gradient energies only  (\ref{eq:vel_s}) as the mean square string velocity estimator, and 
show the means and uncertainties in Fig.~\ref{fig:v_s_Scaling}.

We see that after an initial period of acceleration, 
there is a decreasing trend, approaching what appears to be a constant value at the end of the simulations. 
The maximum velocity is larger for the fields with smaller correlation length in the initial conditions, as is consistent 
with string-like behaviour, where acceleration is proportional to curvature. 
For the case of strings with constant physical width ($s=1$), the RMS velocity is approximately constant during 
the core growth phase, and then approaches an asymptote more slowly than the $s=0$ simulations. 

Figures~\ref{fig:xiScaling} and \ref{fig:v_s_Scaling} show that, independently of the initial field correlation length, all simulations are compatible, \ie all of them give separation and velocity data which are within $1\sigma$ of each other. Moreover, the behaviour of both estimators ($\xi$ and $v$) qualitatively agrees with the standard scaling, showing a tendency towards linear growth in $\xi$ and a constant RMS velocity.

There is a departure from standard scaling in the earlier phases of the simulations, 
which needs to be understood in order to improve 
the estimates of the asymptotic behaviour of $v$ and $\xi$,
or more precisely the asymptotic values of the 
{scaled mean string separation}, 
\ben
x = \xi/t .
\label{x}
\een 
In our previous paper \cite{Hindmarsh:2019csc}, which studied scaling using $\xi_\text{w}$ only, we observed that 
the average slope $ \Delta \xi / \Delta t$ converged 
more quickly to a constant than $\xi / t$. We therefore used the slope of the curve of $\xi$ against $t$ as our estimator 
of the asymptotic value $x_*$. The linear fit can have a significant constant term, which we parameterised in 
terms of the intercept with the time axis, the time offset,  $\tOff$. The value of $\tOff$ has no physical importance, and 
instead parametrises an effect of the initial conditions.

In this paper we make use of RMS velocity data, which gives extra information about the approach to scaling, and 
avoids the need for $\tOff$. We will see that in doing so we improve the accuracy 
of the estimate of $x_*$, while remaining consistent with our previous estimate.

   \begin{figure}[t]
    \centering
    \includegraphics[width=\columnwidth]{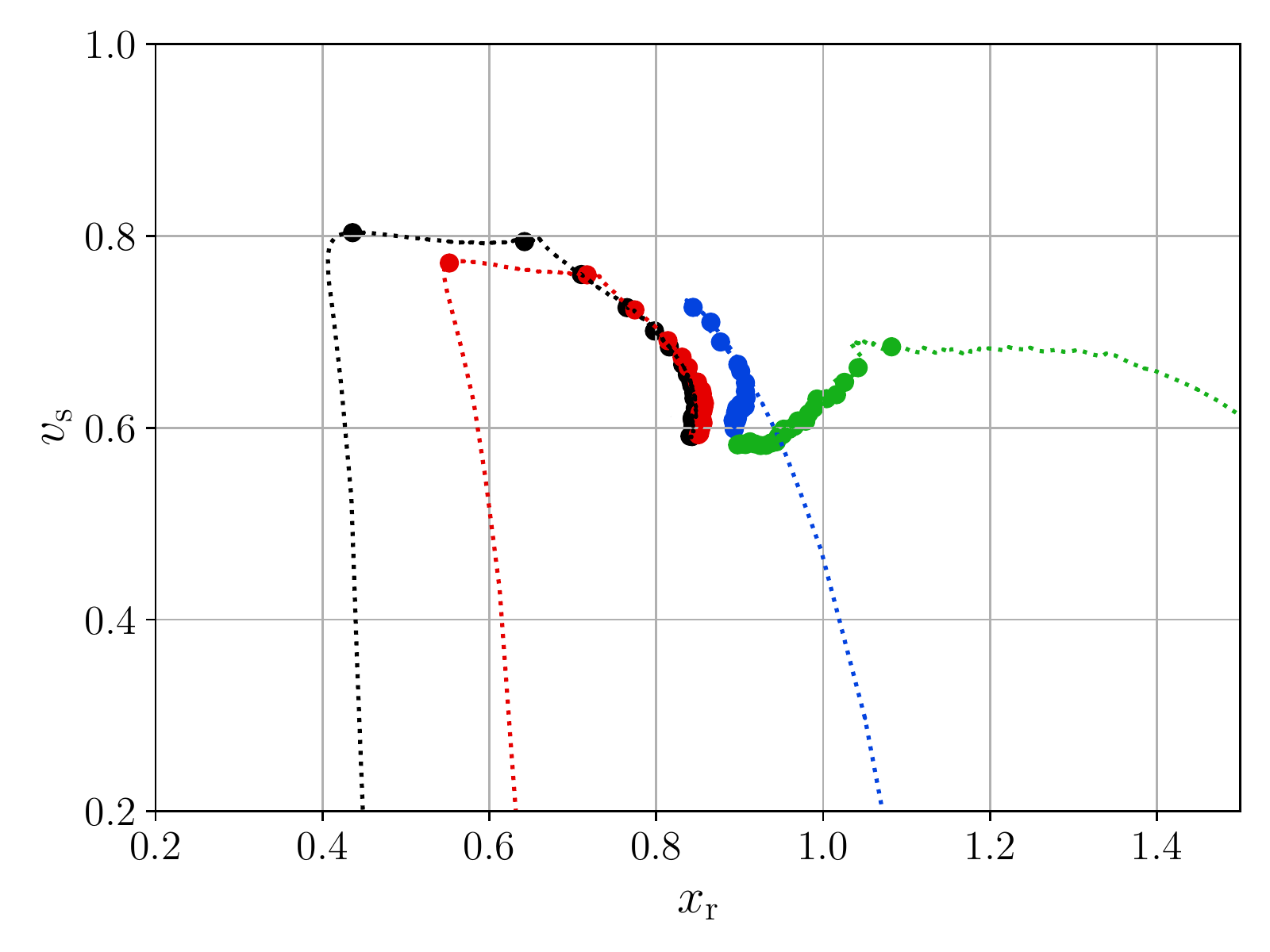}
     \includegraphics[width=\columnwidth]{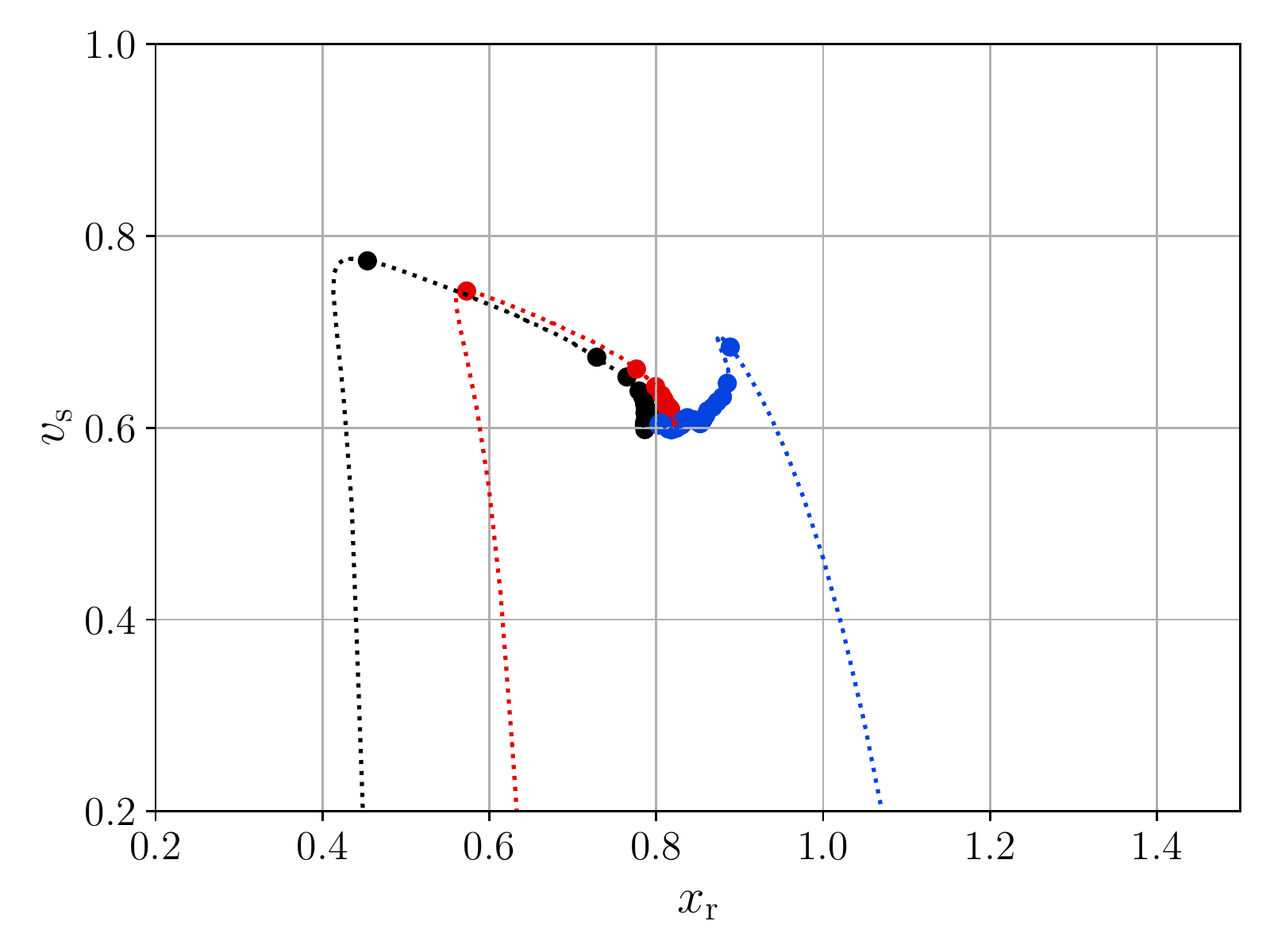}
    \caption{
    Phase space plot for $v_s$ and $\xr$ for $s=1$ (top panel) and $s=0$ (bottom panel), and the same colour scheme correspond to different initial correlation lengths, as the previous figure. 
    Larger dots are plotted every 20 cosmic time units, starting at cosmic time  $t\eta=4$. 
    As time increases, all curves spiral in towards an apparent fixed point, 
    analysed in Section \ref{sec:fits}. 
 \label{fig:phaseplane}}
 \end{figure}

In Fig.~\ref{fig:phaseplane} we show the evolution of the network in the phase space $(x,v)$, where the {scaled mean string separation} is measured with the rest frame string length $\xr$ and the RMS velocity is measured using the scalar field energies $v_\text{s}$. 

The phase space representation shows clearly the different regimes in which the network evolves. At the end of the diffusion period 
the strings are accelerated under their curvature, and the RMS velocity increases rapidly while 
the inter-string distance remains nearly constant. 
For $s=1$ simulations, the diffusive evolution is followed by the core growth period, which is part of the preparation of the 
initial conditions. During the core growth period, velocities remain approximately constant.  The scaled mean string separation, however, changes, and it changes differently for different initial correlation lengths: For correlation lengths $\IniCorLen\eta=5,\ 10$, the scaled mean string separation grows, whereas for correlation lengths  $\IniCorLen\eta=20,\ 40$ it decreases. 
Finally, when the physical equations of motion 
 (\ref{eq:eom}) are being solved, the system starts to spiral towards an apparent common fixed point for all simulations. 
 Estimating the position of the fixed point and hence the asymptotic values of $x$ and $v$ is the subject of the next two sections.

It is interesting to note the qualitative difference in the velocity evolution between the core growth era 
and the physical equations of motion.  In the core growth era the velocity remains constant after the initial 
acceleration: this constant depends on the initial conditions.  It is only after the physical evolution sets in that 
the velocity starts evolving towards its asymptote. 
Note that the core growth era corresponds to evolution with the fixed scale hierarchy $\ms/H$
explored in Ref.~\cite{Klaer:2019fxc}.  We will discuss this observation in the final section.

The initial conditions of the field and the time at which they are set determine 
the simulation's starting point in the phase space. Figure~\ref{fig:phaseplane} shows that varying the initial field correlation length one can choose whether to start on the left hand side or on the right hand side of the hypothetical fixed points, corresponding to strings being either above or below their scaling density. As mentioned before, the time offset $\tOff$ depends on the the initial condition, and for approaches to the fixed point from the right corresponds to $\tOff<0$.

%%%%%%%%%%%%%%%%%%%%%%%%%%%%%%%%%%%%%%%%%%%%%%

\section{Phase space analysis with the VOS model}

In this section we model 
our results as a dynamical system. 
The model best adapted to a network of strings is the velocity-dependent one-scale (VOS) model 
\cite{Kibble:1984hp,Martins:1996jp,Martins:2018dqg}. 
This class of models assumes a statistical distribution of string configurations and velocities which 
has a universal form, parametrised by the string separation $\xi$ (or equivalently the length $\ell$ in a volume $\vol$) 
and RMS velocity $\vAv$.  

When applied to Nambu-Goto strings, the VOS model describes ``long'' strings only, that is, either infinite strings, or 
string loops with total length greater than some threshold of order $\xi$.  
In our  simulations, string lengths and velocities  are measured over the 
whole string network, including loops.  
As a string network's total length is dominated by strings winding around 
the simulation's periodic box, the distinction should not be important for a first approximation.

The movement of the long strings results causes a segment of string of length 
$\xi$ to encounter others at a rate of order $v/\xi$. The encounter causes the string to reconnect, 
producing a loop.  If this loop is smaller than $\xi$ 
it radiates and shrinks without further encounters, apart from self-intersections.  
The net result is the loss of energy from the string network into axions and massive scalar 
radiation.

The string motion is a balance between the acceleration caused by the curvature, and 
the Hubble damping.  There can also be damping due to the preferential loss of energy from 
fast-moving segments of string, which encounter others more rapidly: this effect is 
neglected in the simplest models. There is also direct energy loss from long string in the 
form of radiation, which we will not distinguish from energy loss via loops in our modelling.

The equations of motion for a system of Nambu-Goto strings, and 
the assumptions above 
lead to the following dynamical system,
\bea
\frac{d\xi}{dt} &=& H\xi(1+\vi^2) + \half c\vi, \label{eq:xiVOS} \\
\frac{d\vi}{dt} &=& (1-\vi^2)\left(\frac{k}{\xi}-2H\vi\right), \label{eq:vVOS}
\eea
where 
 $H$ is the Hubble parameter. 
The model has two phenomenological parameters, $k$ and $c$. 
The parameter $c$ describes the efficiency of the energy loss mechanism, 
while the parameter $k$ describes
the correlation between the string curvature vector and the velocity. 
For exactly Nambu-Goto strings, $k$ is a function of velocity. 
However, as a first approximation, and because the strings we are studying are not exactly 
Nambu-Goto, we will take $k$ to be a constant.

Using the dimensionless mean string separation variable $x$ (\ref{x}), 
and taking, $H = 1/2t$ as appropriate for a radiation-dominated universe, 
\begin{eqnarray}
t \dot{x}&=& \half x \left(\vi^2 - 1\right) +\frac{c}{2} \vi\\
t \dot{\vi} &=& \left(1-\vi^2\right)\left(\frac{k}{x}- \vi\right)
\end{eqnarray}
This dynamical system  has 
a fixed point in the relevant region $0 \le  x$, $0 \le \vi < 1$,
\ben
\label{e:FixPoi}
x_*=\sqrt{k( c+k)}\qquad\vi_*=\sqrt{\frac{k}{( c+k)}} .
\een
From here, one can express the parameters  $c$ and $k$  in terms of the fixed point values,
\ben
k = x_*v_*, \quad  c =  \frac{x_*}{v_*} ( 1 - v_*^2). 
\label{fp}
\een
Small perturbations $(\delta x, \delta v)$ evolve to the fixed point according to 
\ben
t \frac{d}{dt} \left( \ba{c} \de x \\ \de v \ea \right) = M_* \left( \ba{c} \de x \\ \de v \ea \right) ,
\een
where 
\ben 
M_*= \left( \begin{array}{cc}
\frac{1}{2}(v_*^2-1)& \frac{ x_*}{2v_*}(1+v_*^2)\\
\frac{v_*}{x_*}(v_*^2-1)&v_*^2-1   \end{array} \right)
\een
The eigenvalues of the matrix $\sigma_\pm$ are 
\[
\sigma_\pm= -\frac{3}{4}\kappa\pm\sqrt{\frac{9}{16}\kappa^2-\kappa}
\]
where   $\kappa = 1 - v_*^2$. Since $0<\kappa<1$ the eigenvalues $\sigma_\pm$ are complex, with negative real part. Therefore, the fixed point is a stable spiral.
 
We plot flows in the phase diagram predicted by the VOS model in Fig.~\ref{f:PhaPlaFlo} for the global best fit $(x_*,v_*)$, along with the mean values of selected $(x,v)$ from the simulations. The stable spiral form is clearly visible in the streamlines. 
 In the next section we explain the fitting procedure from which the global best fit $(x_*,v_*)$ was obtained.

%%%%%%%%%%%%%%%%%%%%%%%%%%%%%%%%%%%%%%%%%%%%%%
\section{Fits and asymptotic behaviour}
\label{sec:fits}

In this section we measure the degree at which the evolution dictated by the VOS model presented in the previous section, Eqs.~(\ref{eq:xiVOS}) and (\ref{eq:vVOS}), is compatible with our simulations,  by performing a fitting analysis. We compute the $\chi^2$ value for each set of 
runs with a given $(s,\IniCorLen)$ as 
\ben
\chi^2=\sum_i \frac{(O_i-E_i)^2}{\si_i^2},
\een
where $O_i$ is the observed value, $E_i$ the expected value on the basis of the model, and $\si_i$ the uncertainty in the observed value. 
In our case, the observed values are the time series data $(x,v)$ recorded from our simulations. We use a bootstrapping method to create the time series for a specific $l_{\phi}$. For each case we have four different runs, out of which we create the bootstrapped time series by choosing randomly a value at specific time step $i$. This procedure is performed
four times so that four different bootstrapped time series are created. The observed values and their uncertainties are then obtained by averaging and computing the standard deviation from those bootstrapped realisations. The expected value is the value predicted by the VOS model, as described below. The set of observations is taken in the time range $[t_{\text{fit}}, t_{\text{end}}]$, where the start of the fitting period is $t_\text{fit}\eta=171.4$ (conformal $\ta_\text{fit}\eta = 600$). Note that the $l_{\phi}\eta=40$ case is present only for $s=1$. 

We explore the two dimensional parameter space using a grid of size $100\times100$, with the priors $0.35<k<0.7$ and $0.6<c<1$. These are set by preliminary analysis of a wider parameter space.

\begin{table}[h!]
\renewcommand{\arraystretch}{1.5}
\scalebox{0.95}{

\begin{tabular}{|c|c|c|c|c|c|c|}
\hline
 $s$ & $l_{\phi}\eta$ &  $k$   & $c$  & $x_*$  & $v_*$   \\ \hline
   \multirow{3}{*}{0} & 5 & $0.474\pm0.006$ & $0.811\pm0.010$ & $0.780\pm0.009$ & $0.607\pm0.002$\\
  & 10 &  $0.486\pm0.003$ & $0.845\pm0.017$ & $0.805\pm0.007$ & $0.604\pm0.004$\\
  & 20 & $0.497\pm0.010$ & $0.764\pm0.010$ & $0.792\pm0.010$ & $0.628\pm0.005$\\ \hline
  \multicolumn{2}{|c|}{Mean} & $0.487\pm0.013$ & $0.803\pm0.032$ & $0.793\pm0.012$ & $0.615\pm0.012$\\
\hline \hline
    \multirow{4}{*}{1}  & 5 & $0.459\pm0.008$ & $0.829\pm0.016$ & $0.768\pm0.013$ & $0.597\pm0.003$\\
  & 10 & $0.485\pm0.011$ & $0.856\pm0.032$ & $0.806\pm0.021$ & $0.601\pm0.004$\\
  & 20 &  $0.519\pm0.004$ & $0.888\pm0.020$ & $0.854\pm0.006$ & $0.607\pm0.005$\\
  & 40 &  $0.521\pm0.005$ & $0.797\pm0.015$ & $0.829\pm0.006$ & $0.629\pm0.005$\\ \hline
   \multicolumn{2}{|c|}{Mean} & $0.494\pm0.027$ & $0.843\pm0.039$ & $0.814\pm0.037$ & $0.609\pm0.014$\\
 \hline
\end{tabular}}
 \caption{\label{tab:ckxv}
 Inferred best-fit values of model parameters $c$ and $k$, and asymptotic values of $x$ and $v$ for each correlation length in $s=0$ and $s=1$. 
 These values are obtained by fitting a set of 20 bootstrap realisations for each correlation length, 
 using data with $t > 171.4$ (conformal time 600). 
The uncertainties on the global means for $s=0$ and $s=1$ are the standard deviations of the mean values for each correlation length.
 }
\end{table}

The best-fit values of $c$ and $k$, and asymptotic scaled mean string separation and velocity ($x_*$ and $v_*$) 
for each $(s,\IniCorLen)$ can be found in Table~\ref{tab:ckxv}, where fits were taken with $\tFit\eta = 171.4$. 
The final mean values for $s=0,1$ are obtained by averaging over different values of $l_{\phi}$, 
and the quoted uncertainties correspond to the resulting standard deviations. 
The errors obtained by quadrature combination of bootstrap errors were systematically smaller than the standard deviations. 
Mean values for the parameters for other values of $\tFit\eta = 171.4$ are shown in Table~\ref{tab:ckxv_early}.

\begin{figure}[h]
    \centering
    \includegraphics[width=\columnwidth]{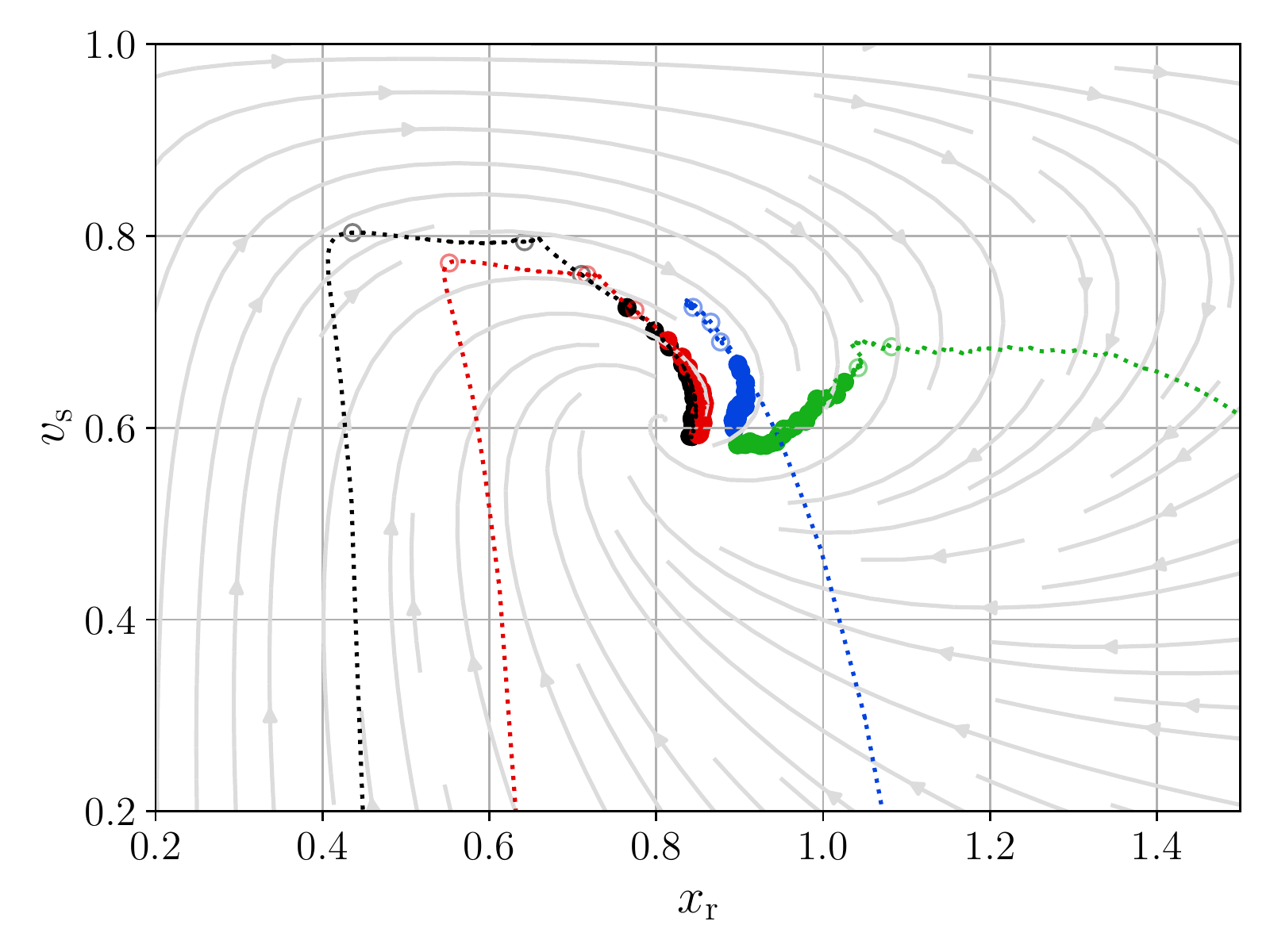}
    \includegraphics[width=\columnwidth]{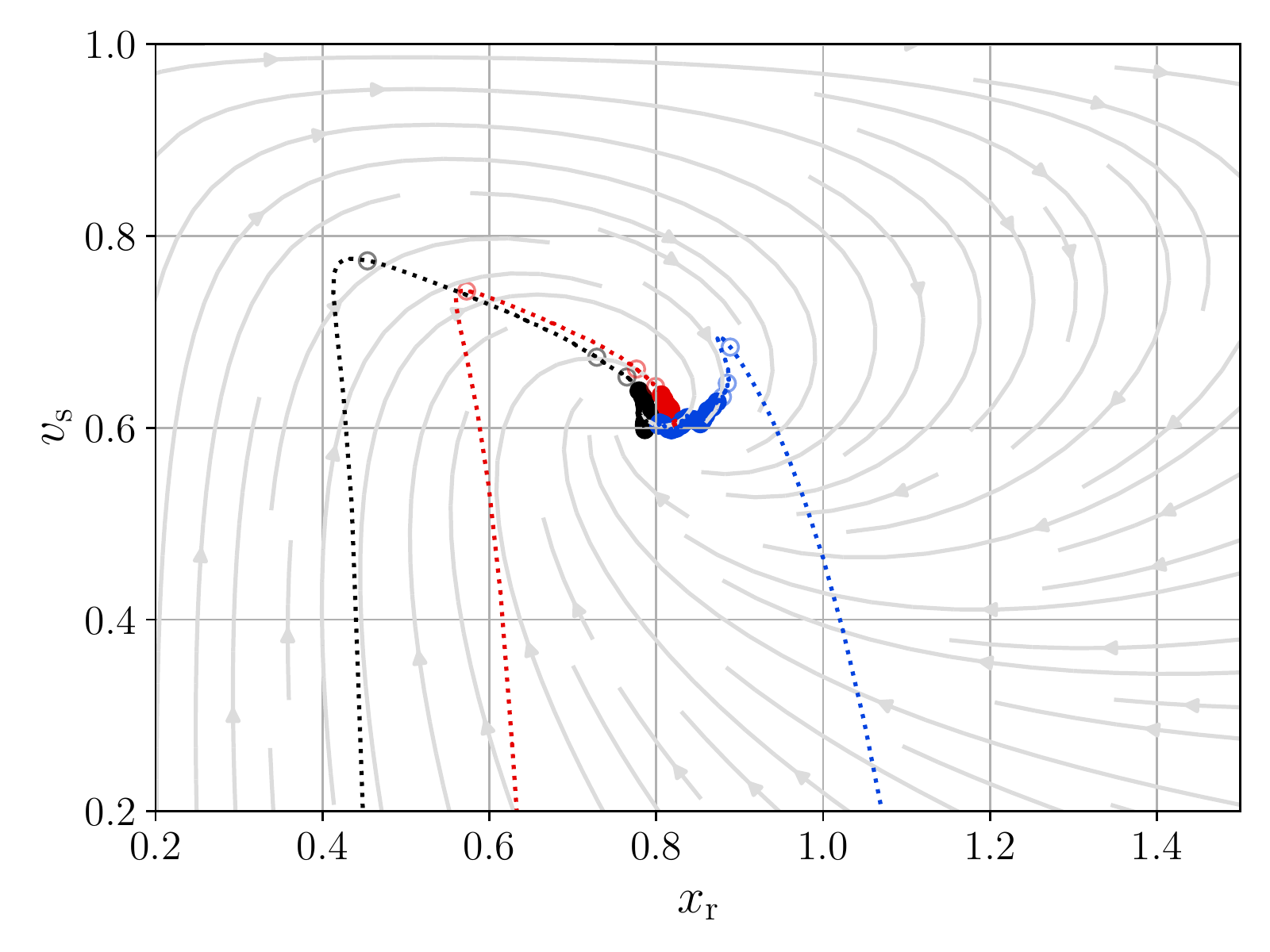}
    \caption{\label{f:PhaPlaFlo}
Phase plane for $s=1$ (top),  $s=0$ (bottom), with stream lines of the best-fit values of the VOS model parameters shown in Table~\ref{tab:ckxv_early}, with $\tFit\eta = 50$. 
 The colour scheme is the same as in previous figures. 
 The larger markers correspond to the same points as in Fig.~\ref{fig:phaseplane}, 
 with empty circles denoting points with $\tFit\eta < 50$.}
 \end{figure}
 
 Figure \ref{f:PhaPlaFlo} shows the evolution of the simulations alongside the streamlines of the VOS dynamical system calculated using the inferred global mean values $(x_*, v_*)$, obtained by fitting with $\tFit\eta = 50$, 
 the earliest time from which we start fitting (see Table~\ref{tab:ckxv_early}).
 It can be seen that after an initial relaxation period, 
 the simulation data follow the spiral-like evolution towards the fixed point.  A key feature is that initial conditions with $x < x_*$ tend to flow to states with  $v > v_*$, 
 and then around to $x> x_*$, corresponding to string networks less dense than scaling. 

\begin{figure}[h]
    \centering
    \includegraphics[width=\columnwidth]{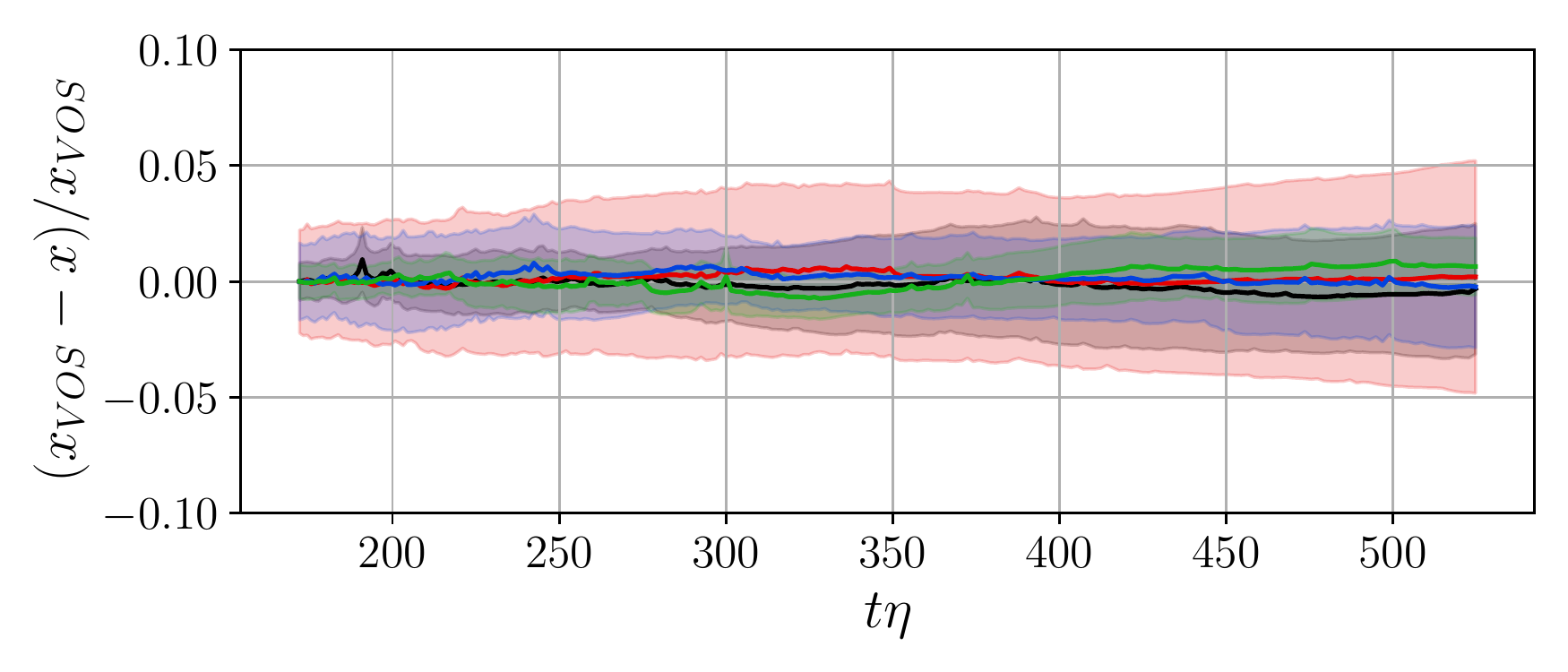}
     \includegraphics[width=\columnwidth]{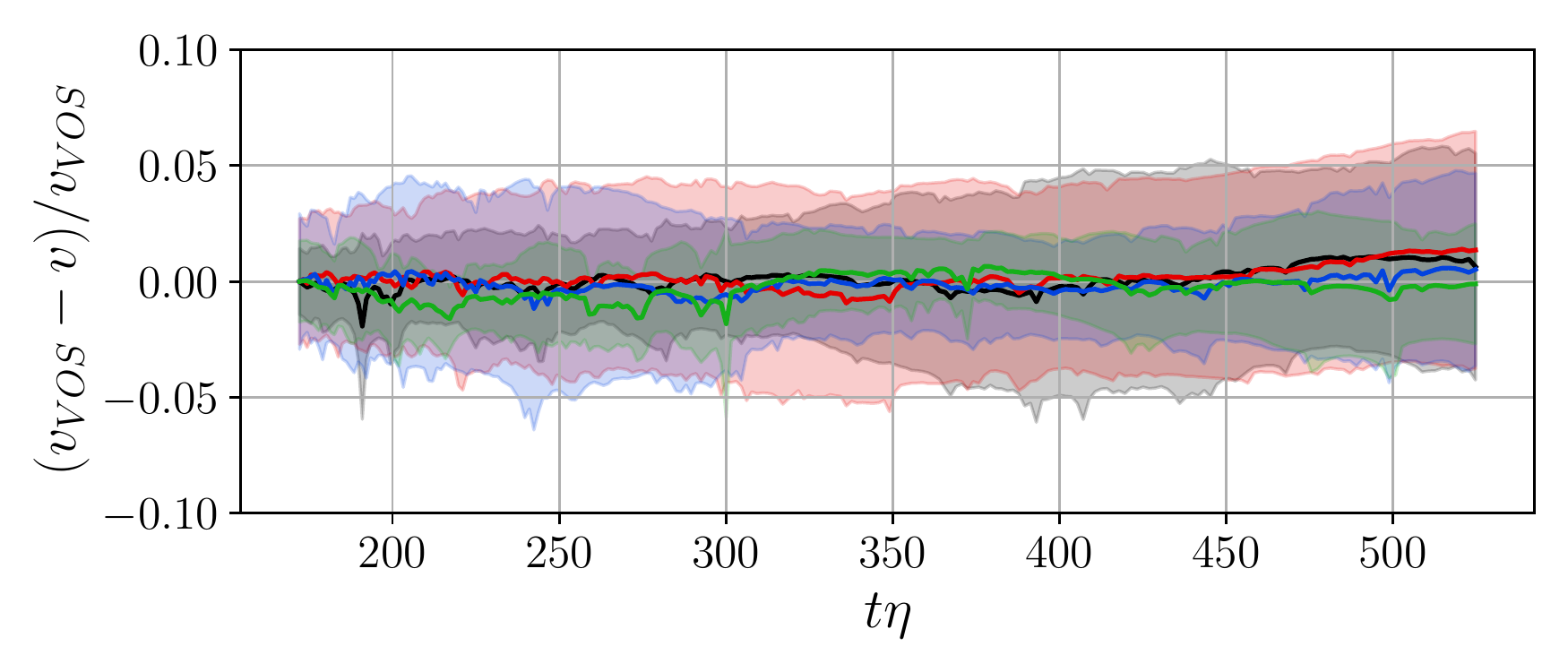}
    \caption{\label{f:Resi}
    Relative difference of the VOS prediction and the simulation data of the dimensionless string separation $x$ and rms velocity $v$. 
    The shaded bands represent errors propagated from the simulations' energy estimators.}
 \end{figure}

For a more quantitative comparison between our simulations and the VOS model, 
we show in 
Figure~\ref{f:Resi} the relative difference between the simulation time series data and the VOS best-fit model for each $(s, \IniCorLen)$, 
where the initial conditions for the integration of the VOS equations are set at $t_{\text{fit}}$. Shaded regions correspond to the uncertainties propagated from simulation estimators. 
It can be observed that the mean relative difference always lies below $5\%$ level of deviation, with zero deviation always within the errors. 

\begin{table}[h!]
\renewcommand{\arraystretch}{1.5}
\scalebox{0.95}{

\begin{tabular}{|c|c|c|c|c|c|}
\hline
 $s$ & $t_{\text{fit}}\eta$ &  $k$   & $c$  & $x_*$  & $v_*$   \\ \hline
 \multirow{10}{*}{0} 
 & 50 & $0.486\pm0.027$ & $0.804\pm0.008$ & $0.793\pm0.030$ & $0.614\pm0.011$\\
 & 100 & $0.487\pm0.018$ & $0.800\pm0.018$ & $0.792\pm0.022$ & $0.615\pm0.007$\\
 & 150 & $0.484\pm0.017$ & $0.804\pm0.034$ & $0.790\pm0.020$ & $0.613\pm0.010$\\
 & 171 & $0.486\pm0.012$ & $0.807\pm0.041$ & $0.792\pm0.013$ & $0.613\pm0.013$\\
 & 233 & $0.478\pm0.011$ & $0.808\pm0.053$ & $0.783\pm0.026$ & $0.610\pm0.010$\\
 & 305 & $0.450\pm0.015$ & $0.789\pm0.077$ & $0.746\pm0.025$ & $0.604\pm0.022$\\
 & 386 & $0.450\pm0.036$ & $0.745\pm0.115$ & $0.732\pm0.029$ & $0.616\pm0.040$\\
\hline
 \multirow{7}{*}{1} 
 & 50 & $0.493\pm0.057$ & $0.840\pm0.013$ & $0.810\pm0.062$ & $0.607\pm0.024$\\
 & 100 & $0.493\pm0.043$ & $0.841\pm0.015$ & $0.811\pm0.045$ & $0.607\pm0.020$\\
 & 150 & $0.498\pm0.031$ & $0.837\pm0.021$ & $0.815\pm0.036$ & $0.611\pm0.012$\\
 & 171 & $0.496\pm0.030$ & $0.843\pm0.039$ & $0.814\pm0.037$ & $0.609\pm0.014$\\
 & 233 & $0.492\pm0.028$ & $0.835\pm0.046$ & $0.808\pm0.037$ & $0.609\pm0.013$\\
 & 305 & $0.484\pm0.028$ & $0.854\pm0.083$ & $0.804\pm0.032$ & $0.602\pm0.026$\\
 & 386 & $0.483\pm0.060$ & $0.846\pm0.102$ & $0.799\pm0.050$ & $0.604\pm0.045$\\
 \hline
\end{tabular}}
 \caption{\label{tab:ckxv_early}
 Global mean values of model parameters $c$ and $k$, and asymptotic values of $x$ and $v$, 
 with different start times for the fit $\tFit$. 
 }
\end{table}

Table~\ref{tab:ckxv_early} contains the global mean values obtained applying the same bootstrap procedure as in the previous analysis. 
The last four fit start times are those used for 
the linear fitting procedure carried out in Ref.~\cite{Hindmarsh:2019csc}. 
A remarkably good agreement is obtained in the global means when comparing early and late fits. Earlier fits have a smaller scatter in $c$, but a larger scatter in $k$. 
This can be related to the spread of velocities in the initial conditions
The spread is minimised for the intermediate times $\tFit = 150, 171$ and $233$. 
We quote results from $\tFit = 171$, which is also the earliest fitting time from our 
previous paper \cite{Hindmarsh:2019csc}, meaning that a direct comparison 
of the methods can be made.

The simplest VOS model therefore gives a good quantitative description of the joint evolution of the string separation and RMS velocity.  
We have experimented with fitting for additional parameters 
$q$ and $d$ (with $\beta = 1$ and $r = 1$) in the VOS model presented in Ref.~\cite{Correia:2019bdl}. 
but our preliminary analysis shows that the preferred values for these 
additional parameters are compatible with zero.

\begin{figure}[h]
    \centering
    \includegraphics[width=\columnwidth]{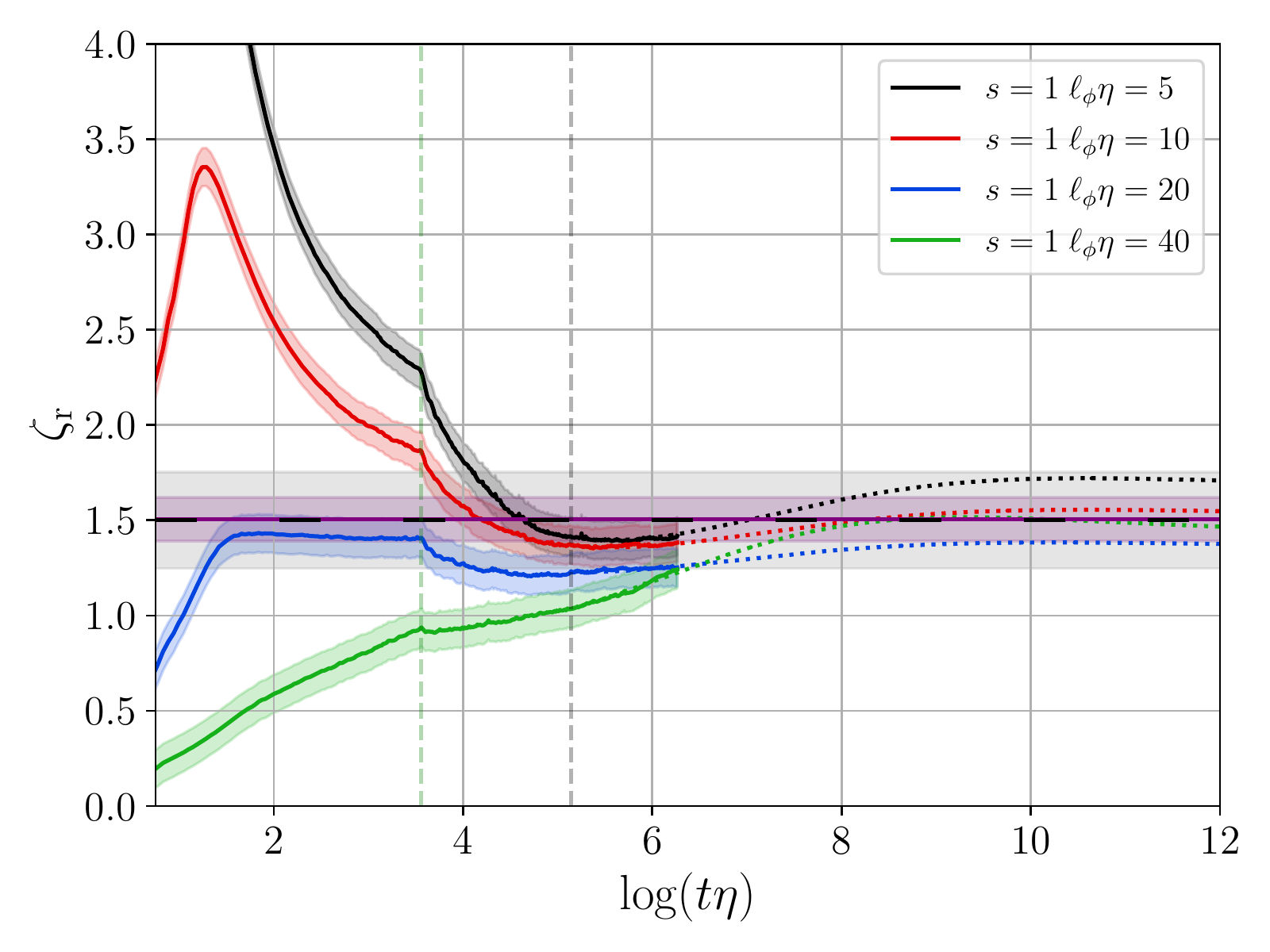}
     \includegraphics[width=\columnwidth]{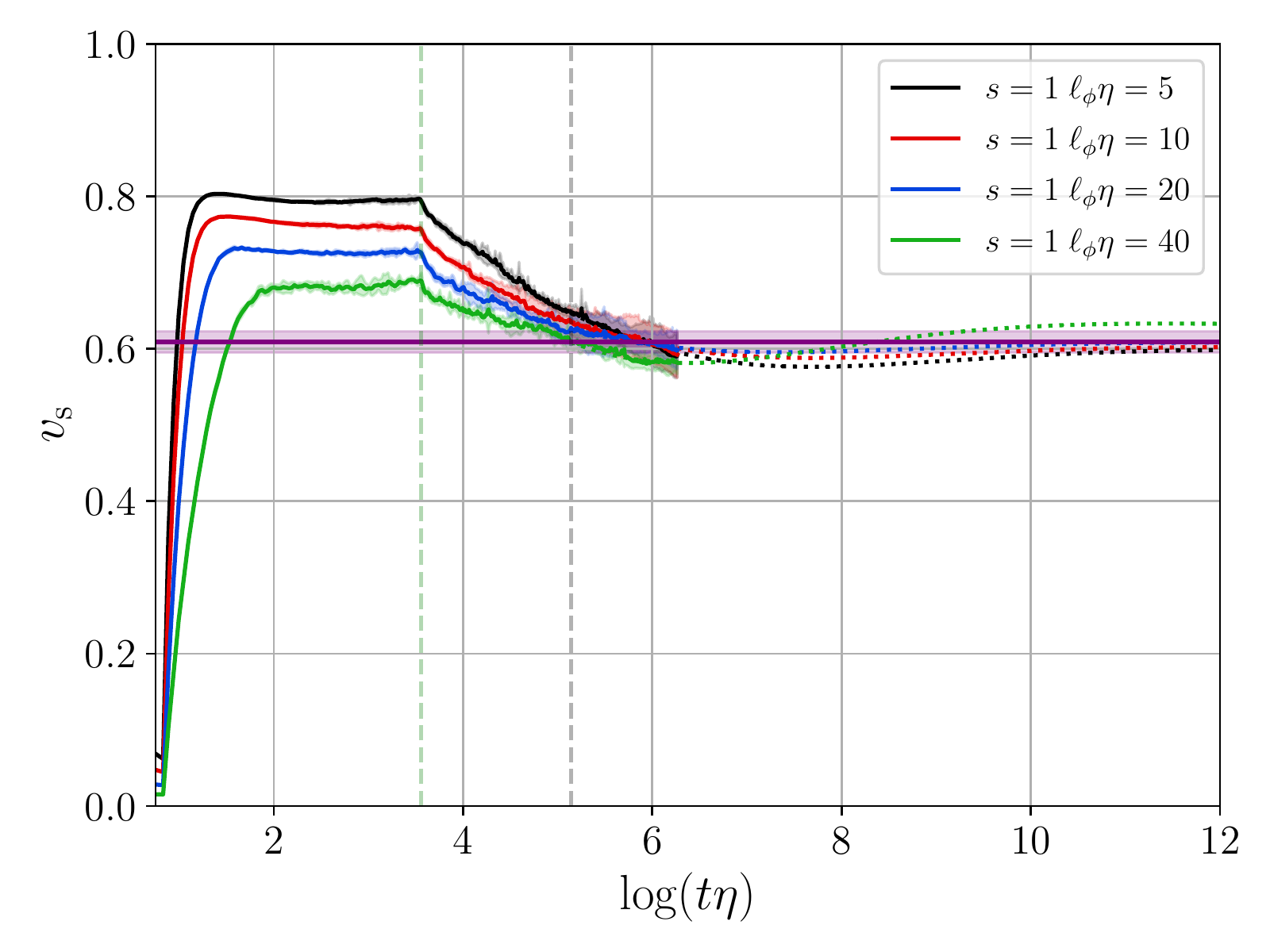}
    \caption{\label{f:Log}
String network evolution expressed in terms of the rest-frame length density parameter $\zeta$ and RMS velocity $v$, plotted against $\log(t\eta)$, with VOS models (dotted line) that correspond to best fit values for each $\IniCorLen$ shown in Table~\ref{tab:ckxv}. The prediction of the VOS model is plotted from $t_{\rm fit}$ on. The horizontal dashed black line and grey band in the top panel show the mean and uncertainly obtained from our previous analysis \cite{Hindmarsh:2019csc}, translated from the universe-frame length used in that paper by multiplying  by $(1 - v_*^2)^{-1/2}$.  
The solid purple line and shaded purple bands are the mean and uncertainly obtained from the analysis in this work.
The vertical green dashed line marks the end of the core growth phase, and 
the fit start time $\tFit$ is indicated by the vertical grey dashed line.
}
 \end{figure}

It is also interesting to study the network evolution in terms of the length density parameter $\ze$ (\ref{e:ZetDef}), and 
by plotting against the logarithm of time one can emphasise the earlier times when the network is 
further away from scaling. Fig.~\ref{f:Log} shows our $s=1$ rest-frame length data plotted this way, along with 
the best-fit VOS models for each correlation length, and their extrapolation to larger values of time. The asymptotic $\zeta_{\text{r},*}$ obtained from the overall mean values of fit parameters in Table~\ref{tab:ckxv} is also depicted, for which we obtain $\zeta_{\text{r},*} = 1.50 \pm 0.11$. The central value is shown as solid purple line and its corresponding errors in shaded purple bands. Note that all simulations approach the asymptotic $\ze$ from below, and are still slowly 
increasing at the end of the simulation, but within $20$\% of its asymptotic value.    
The increase is most noticeable for simulations which start very underdense, and therefore 
have further to evolve to reach scaling.

 We also include the value of the asymptotic length density parameter reported in our previous paper \cite{Hindmarsh:2019csc} in dashed black with its corresponding uncertainty in grey. In \cite{Hindmarsh:2019csc} 
 the universe-frame string length $\SlenWin$ was used as the measure of the string length, 
and the value of $x_*$ estimated by linear fits to $\xi_\text{w}$ against $t$, from which we obtained an estimate of 
the asymptotic value 
$\zeta_{*} = 1.19 \pm 0.20$. In this work we use the rest-frame length estimator, the corresponding string density parameter is (\ref{e:LenWinRes}), for which $\zeta_{\text{r},*} =  \zeta_{*}  (1 - v_*^2)^{-1/2}$, and gives $\zeta_{\text{r},*} = 1.51\pm 0.25$.
 As the figure shows, the agreement with our previous result is very good. 

%%%%%%%%%%%%%%%%%%%%%%%%%%%%%%%%%%%%%%%%%%%%%%
\section{Discussion and conclusions}
\label{sec:Con}

In this paper we have studied axion string networks in the radiation era by measuring   the root-mean-square velocity of the strings $\vAv$ and the mean string separation $\xi$.  The strings are modelled by a scalar field with two real components with a spontaneously broken O(2) symmetry, simulated in periodic cubic lattices.
Performing a phase space analysis in the variables $x = \xi/t$ and $\vAv$, we find good evidence for the existence of a fixed point, which shows that the system reaches a scaling regime. These prompted us to continue the analysis  in the framework of the velocity-dependent  one-scale (VOS) model. 

The VOS model assumes a statistical distribution of string positions and velocities which can be adequately described 
by two parameters mentioned above.
By assuming that the strings follow approximately Nambu-Goto trajectories, that they reconnect with a fixed high probability when they 
cross, and that the loops so formed annihilate quickly, the VOS model reduces the network evolution to a simple dynamical 
system, a pair of first order non-linear ordinary differential equations.  The equations have 
a fixed point $(x_*,\vAv_*)$, which describes a scaling network, that is, one whose mean separation increases linearly with time, and 
with constant RMS velocity.

We have fitted the results of a set of numerical simulations in the radiation era to a two-parameter VOS model, 
with initially random fields with several different initial correlations lengths $\IniCorLen$, 
using both  the true physical field equations and the PRS approximation. In terms of the core growth parameter $s$, 
the comoving string width behaves as $\wss = {\ws}{a^{-s}} $.  In this paper we have used  $s=1$, which corresponds to the true physical case, and $s=0$ which corresponds to a string with constant comoving width. 

We find that the two-parameter VOS model gives a good qualitative and quantitative description 
of the network evolution, with parameters given in Table \ref{tab:ckxv}.  

Qualitatively, 
the initial acceleration of the string network results in a RMS velocity which overshoots the fixed point $v_*$, as the Hubble length in our initial conditions is larger than the string separation, meaning that the dynamical system is underdamped.  
The higher velocity results in more rapid loop formation, and hence an 
increase in the mean string separation.  This decreases the acceleration, and hence the RMS velocity.  The net result is a curved  
approach to the fixed point in the $(x,v)$ plane, clearly visible in Fig.~\ref{f:PhaPlaFlo}.  

In assessing the quantitative success of the two-parameter VOS model, we observe that the residuals to the fits points in Fig.~\ref{f:Resi} are consistent with zero, and that the 
fixed points given in Table \ref{tab:ckxv} for differential initial correlation lengths are remarkably similar. 
The fluctuations between the fixed point estimates are slightly larger than the bootstrap fitting errors would predict, 
which suggests that the model could be tuned slightly, or that the fitting errors have been underestimated. A preliminary investigation shows that the more complex 
model of Ref.~\cite{Correia:2019bdl} does not improve the fit. 
A  more thorough exploration of VOS models and a more accurate estimate of the fixed point could be obtained with a wider range of initial correlation lengths and initial times.

Translating the  values of Table \ref{tab:ckxv} to the universe-frame length density parameter $\ze$, estimated in our previous paper by linear fitting, 
we find
\bea
\ze_* &=&  \zzMeansOneNew \pm  \zzErrsOneNew \; (s=1), \\
\ze_* &=& \zzMeansZeroNew \pm  \zzErrsZeroNew \; (s=0). 
\eea
These values are consistent with our previous determination, with an improved accuracy arising from the 
joint fit with the velocity data in the context of the VOS model.

An important consequence of the description in terms of a dynamical system is that the approach to the fixed point is determined by a pair of complex exponents $\si_\pm$, whose real part is $-\frac{3}{4}( 1- v_*^2) \simeq -0.47$. Hence, even when close to the fixed point, the approach can be rather slow. 

If the initial string separation $\xi_\text{i}$ is chosen far away from its scaling value $x_* t_\text{i}$, 
it may not get within $1\si$ of its scaling value (as determined by the VOS model) by the end of the simulation, 
which has to be chosen as $L/2$ for a box of side $L$ in systems like this one with degrees of freedom propagating at the speed of light.

This is particularly noticeable for initial conditions which are very underdense, i.e.~with $x_\text{i} \gg x_*$. 
When the length density parameter $\ze$ is plotted against the logarithm of cosmic time $\log(t\eta)$, 
one sees a slow drift up towards the fixed point value.
Other groups have also noticed this feature of underdense initial conditions 
\cite{Gorghetto:2018myk,Kawasaki:2018bzv,Vaquero:2018tib,Buschmann:2019icd,Klaer:2019fxc,Gorghetto:2020qws}.
As we have explained elsewhere \cite{Hindmarsh:2019csc} this does not signal a 
breakdown of the standard scaling picture. 
Our analysis in the framework of the VOS model shows that slow approaches to the fixed 
point from values of $\zeta$ less than its fixed point value are to be expected, 
and indeed, 
nearly all simulations to date have final values of $\ze$ less than our 
estimated fixed point.  

If the slow upward drift in $\ze$ were an asymptotic feature of axion string networks, one would expect to see 
final values of $\ze$ significantly above the fixed-point value in the largest simulations. 
However, the maximum value of $\ze$ obtained in the most recent (and therefore largest) simulations 
are nearly all below 
value of $\ze_*$ computed in this work.
These values are (all of them measured in the universe frame):  
$\ze\simeq0.9$ in the physical case and $\ze\simeq1.2$ using the PRS approximation  in \cite{Gorghetto:2020qws},  $\ze\simeq 1.1$ in the physical case in \cite{Klaer:2019fxc} and  $\ze\simeq1.4$ using the PRS approximation in \cite{Vaquero:2018tib}. In \cite{Kawasaki:2018bzv} the authors show the physical evolution of $\ze$ for three different ratios of the Hubble scale to the string width at the time of the PQ phase transition, giving $\ze\simeq 1.3$, $\ze\simeq 1.1$ and $\ze\simeq 0.9$. In \cite{Buschmann:2019icd}, 
the quoted value $\ze\simeq4$ is not consistent with other groups, but  the authors of that work caution that the method to detect strings they use gives only a rough estimate of $\ze$. 
They also suggest that a better method will render their results comparable with the ones in \cite{Gorghetto:2020qws}, and therefore, also with other groups.
 
In summary, our data and fits already show that the straightforward and physically 
motivated picture provided by the simplest VOS model 
provides a good description of the evolution of the string network 
consistent with the standard scaling picture, and an asymptotically constant 
dimensionless length density and RMS velocity. 
This gives confidence that our results can be extrapolated over the many orders of magnitude 
required for predictions of the axion number density.
As pointed out in Ref.~\cite{Hindmarsh:2019csc}, predictions from a scaling string 
network will be around 50\% higher than recent estimates \cite{Klaer:2017ond}.  

We also comment on a suggestion that simulations 
with growing comoving string width shed light on the asymptotic behaviour of axion string 
networks \cite{Klaer:2019fxc}. 
If the string width grows in 
proportion to the horizon ($\wss \propto \tau$), the field equations have a scale symmetry, 
and it can be argued that a fixed point must exist. 
We use this growth in width in the core growth phase of our $s=1$ simulations, 
where it can can be seen that this phase characterised by a constant RMS 
velocity, but not a consistent one.  A smaller initial correlation length gives a larger RMS velocity in the core growth phase (see the top panels in Figs.~\ref{f:Log} and \ref{fig:v_s_Scaling}). 
This is understandable in that the initial acceleration is proportional to the curvature.
However, as the mean string separation increases, the RMS velocity of the strings does not decrease, as 
would be expected from the decrease in the average acceleration.  
While it seems that $x$ is evolving towards a value $x_* \simeq 0.8$, there is little sign of a 
definite value of the RMS velocity.  
It seems therefore that networks in the core growth phase can be used 
to estimate the fixed point in $x$, but not the RMS velocity.  

As a final remark, we note that the asymptotic scaling behaviour presented here can also be applicable to generic global string networks, 
such as those in axion models beyond the canonical QCD scenario \cite{Svrcek:2006yi}. A general observational consequence of scaling in a system of topological defects is a scale-invariant gravitational wave power spectrum \cite{Figueroa:2012kw,Figueroa:2020lvo} during radiation domination. 
This suggests that recent claim that axion string networks 
produce a tilted gravitational wave spectrum  \cite{Gorghetto:2021fsn}, based on an 
assumed logarithmic growth in the string density throughout the radiation era, should be revisited. 

\begin{acknowledgments}
We are grateful to Daniel Cutting and Daniel G. Figueroa for comments on the draft manuscript.
MH (ORCID ID 0000-0002-9307-437X) acknowledges support from the Science and Technology Facilities Council (grant number ST/L000504/1)
and the Academy of Finland (grant number 333609). JL (ORCID ID 0000-0002-1198-3191) and JU (ORCID ID 0000-0002-4221-2859) acknowledge support from Eusko Jaurlaritza (IT-979-16) and PGC2018-094626-B-C21 (MCIU/AEl/FEDER,UE). ALE (ORCID ID 0000-0002-1696-3579) is supported by the National Science Foundation grant PHY-1820872. 
ALE is grateful to the Early Universe Cosmology group of the University of the Basque Country for their generous hospitality and useful discussions.  
This research was supported by the Munich Institute for Astro- and Particle Physics (MIAPP) which is funded by the Deutsche Forschungsgemeinschaft (DFG, German Research Foundation) under Germany's Excellence Strategy – EXC-2094 – 390783311.
This work has been possible thanks to the computational resources on the STFC DiRAC HPC facility obtained under the dp116 project. Our simulations also made use of facilities at the i2Basque academic network and CSC Finland.
\end{acknowledgments}

\bibliography{axion} % Produces the bibliography via BibTeX.

\begin{thebibliography}{57}%
\makeatletter
\providecommand \@ifxundefined [1]{%
 \@ifx{#1\undefined}
}%
\providecommand \@ifnum [1]{%
 \ifnum #1\expandafter \@firstoftwo
 \else \expandafter \@secondoftwo
 \fi
}%
\providecommand \@ifx [1]{%
 \ifx #1\expandafter \@firstoftwo
 \else \expandafter \@secondoftwo
 \fi
}%
\providecommand \natexlab [1]{#1}%
\providecommand \enquote  [1]{``#1''}%
\providecommand \bibnamefont  [1]{#1}%
\providecommand \bibfnamefont [1]{#1}%
\providecommand \citenamefont [1]{#1}%
\providecommand \href@noop [0]{\@secondoftwo}%
\providecommand \href [0]{\begingroup \@sanitize@url \@href}%
\providecommand \@href[1]{\@@startlink{#1}\@@href}%
\providecommand \@@href[1]{\endgroup#1\@@endlink}%
\providecommand \@sanitize@url [0]{\catcode `\\12\catcode `\$12\catcode
  `\&12\catcode `\#12\catcode `\^12\catcode `\_12\catcode `\%12\relax}%
\providecommand \@@startlink[1]{}%
\providecommand \@@endlink[0]{}%
\providecommand \url  [0]{\begingroup\@sanitize@url \@url }%
\providecommand \@url [1]{\endgroup\@href {#1}{\urlprefix }}%
\providecommand \urlprefix  [0]{URL }%
\providecommand \Eprint [0]{\href }%
\providecommand \doibase [0]{http://dx.doi.org/}%
\providecommand \selectlanguage [0]{\@gobble}%
\providecommand \bibinfo  [0]{\@secondoftwo}%
\providecommand \bibfield  [0]{\@secondoftwo}%
\providecommand \translation [1]{[#1]}%
\providecommand \BibitemOpen [0]{}%
\providecommand \bibitemStop [0]{}%
\providecommand \bibitemNoStop [0]{.\EOS\space}%
\providecommand \EOS [0]{\spacefactor3000\relax}%
\providecommand \BibitemShut  [1]{\csname bibitem#1\endcsname}%
\let\auto@bib@innerbib\@empty
%</preamble>
\bibitem [{\citenamefont {Weinberg}(1978)}]{Weinberg:1977ma}%
  \BibitemOpen
  \bibfield  {author} {\bibinfo {author} {\bibfnamefont {S.}~\bibnamefont
  {Weinberg}},\ }\href {\doibase 10.1103/PhysRevLett.40.223} {\bibfield
  {journal} {\bibinfo  {journal} {Phys. Rev. Lett.}\ }\textbf {\bibinfo
  {volume} {40}},\ \bibinfo {pages} {223} (\bibinfo {year} {1978})}\BibitemShut
  {NoStop}%
%%CITATION = PRLTA,40,223;%%
\bibitem [{\citenamefont {Wilczek}(1978)}]{Wilczek:1977pj}%
  \BibitemOpen
  \bibfield  {author} {\bibinfo {author} {\bibfnamefont {F.}~\bibnamefont
  {Wilczek}},\ }\href {\doibase 10.1103/PhysRevLett.40.279} {\bibfield
  {journal} {\bibinfo  {journal} {Phys. Rev. Lett.}\ }\textbf {\bibinfo
  {volume} {40}},\ \bibinfo {pages} {279} (\bibinfo {year} {1978})}\BibitemShut
  {NoStop}%
%%CITATION = PRLTA,40,279;%%
\bibitem [{\citenamefont {Peccei}\ and\ \citenamefont
  {Quinn}(1977{\natexlab{a}})}]{Peccei:1977hh}%
  \BibitemOpen
  \bibfield  {author} {\bibinfo {author} {\bibfnamefont {R.~D.}\ \bibnamefont
  {Peccei}}\ and\ \bibinfo {author} {\bibfnamefont {H.~R.}\ \bibnamefont
  {Quinn}},\ }\href {\doibase 10.1103/PhysRevLett.38.1440} {\bibfield
  {journal} {\bibinfo  {journal} {Phys. Rev. Lett.}\ }\textbf {\bibinfo
  {volume} {38}},\ \bibinfo {pages} {1440} (\bibinfo {year}
  {1977}{\natexlab{a}})}\BibitemShut {NoStop}%
%%CITATION = PRLTA,38,1440;%%
\bibitem [{\citenamefont {Peccei}\ and\ \citenamefont
  {Quinn}(1977{\natexlab{b}})}]{Peccei:1977ur}%
  \BibitemOpen
  \bibfield  {author} {\bibinfo {author} {\bibfnamefont {R.~D.}\ \bibnamefont
  {Peccei}}\ and\ \bibinfo {author} {\bibfnamefont {H.~R.}\ \bibnamefont
  {Quinn}},\ }\href {\doibase 10.1103/PhysRevD.16.1791} {\bibfield  {journal}
  {\bibinfo  {journal} {Phys. Rev.}\ }\textbf {\bibinfo {volume} {D16}},\
  \bibinfo {pages} {1791} (\bibinfo {year} {1977}{\natexlab{b}})}\BibitemShut
  {NoStop}%
%%CITATION = PHRVA,D16,1791;%%
\bibitem [{\citenamefont {Kim}(1979)}]{Kim:1979if}%
  \BibitemOpen
  \bibfield  {author} {\bibinfo {author} {\bibfnamefont {J.~E.}\ \bibnamefont
  {Kim}},\ }\href {\doibase 10.1103/PhysRevLett.43.103} {\bibfield  {journal}
  {\bibinfo  {journal} {Phys. Rev. Lett.}\ }\textbf {\bibinfo {volume} {43}},\
  \bibinfo {pages} {103} (\bibinfo {year} {1979})}\BibitemShut {NoStop}%
%%CITATION = PRLTA,43,103;%%
\bibitem [{\citenamefont {Shifman}\ \emph {et~al.}(1980)\citenamefont
  {Shifman}, \citenamefont {Vainshtein},\ and\ \citenamefont
  {Zakharov}}]{Shifman:1979if}%
  \BibitemOpen
  \bibfield  {author} {\bibinfo {author} {\bibfnamefont {M.~A.}\ \bibnamefont
  {Shifman}}, \bibinfo {author} {\bibfnamefont {A.~I.}\ \bibnamefont
  {Vainshtein}}, \ and\ \bibinfo {author} {\bibfnamefont {V.~I.}\ \bibnamefont
  {Zakharov}},\ }\href {\doibase 10.1016/0550-3213(80)90209-6} {\bibfield
  {journal} {\bibinfo  {journal} {Nucl. Phys.}\ }\textbf {\bibinfo {volume}
  {B166}},\ \bibinfo {pages} {493} (\bibinfo {year} {1980})}\BibitemShut
  {NoStop}%
%%CITATION = NUPHA,B166,493;%%
\bibitem [{\citenamefont {Zhitnitsky}(1980)}]{Zhitnitsky:1980tq}%
  \BibitemOpen
  \bibfield  {author} {\bibinfo {author} {\bibfnamefont {A.~R.}\ \bibnamefont
  {Zhitnitsky}},\ }\href@noop {} {\bibfield  {journal} {\bibinfo  {journal}
  {Sov. J. Nucl. Phys.}\ }\textbf {\bibinfo {volume} {31}},\ \bibinfo {pages}
  {260} (\bibinfo {year} {1980})},\ \bibinfo {note} {[Yad.
  Fiz.31,497(1980)]}\BibitemShut {NoStop}%
%%CITATION = SJNCA,31,260;%%
\bibitem [{\citenamefont {Dine}\ \emph {et~al.}(1981)\citenamefont {Dine},
  \citenamefont {Fischler},\ and\ \citenamefont {Srednicki}}]{Dine:1981rt}%
  \BibitemOpen
  \bibfield  {author} {\bibinfo {author} {\bibfnamefont {M.}~\bibnamefont
  {Dine}}, \bibinfo {author} {\bibfnamefont {W.}~\bibnamefont {Fischler}}, \
  and\ \bibinfo {author} {\bibfnamefont {M.}~\bibnamefont {Srednicki}},\ }\href
  {\doibase 10.1016/0370-2693(81)90590-6} {\bibfield  {journal} {\bibinfo
  {journal} {Phys. Lett.}\ }\textbf {\bibinfo {volume} {104B}},\ \bibinfo
  {pages} {199} (\bibinfo {year} {1981})}\BibitemShut {NoStop}%
%%CITATION = PHLTA,104B,199;%%
\bibitem [{\citenamefont {Preskill}\ \emph {et~al.}(1983)\citenamefont
  {Preskill}, \citenamefont {Wise},\ and\ \citenamefont
  {Wilczek}}]{Preskill:1982cy}%
  \BibitemOpen
  \bibfield  {author} {\bibinfo {author} {\bibfnamefont {J.}~\bibnamefont
  {Preskill}}, \bibinfo {author} {\bibfnamefont {M.~B.}\ \bibnamefont {Wise}},
  \ and\ \bibinfo {author} {\bibfnamefont {F.}~\bibnamefont {Wilczek}},\ }\href
  {\doibase 10.1016/0370-2693(83)90637-8} {\bibfield  {journal} {\bibinfo
  {journal} {Phys. Lett.}\ }\textbf {\bibinfo {volume} {B120}},\ \bibinfo
  {pages} {127} (\bibinfo {year} {1983})}\BibitemShut {NoStop}%
%%CITATION = PHLTA,B120,127;%%
\bibitem [{\citenamefont {Abbott}\ and\ \citenamefont
  {Sikivie}(1983)}]{Abbott:1982af}%
  \BibitemOpen
  \bibfield  {author} {\bibinfo {author} {\bibfnamefont {L.~F.}\ \bibnamefont
  {Abbott}}\ and\ \bibinfo {author} {\bibfnamefont {P.}~\bibnamefont
  {Sikivie}},\ }\href {\doibase 10.1016/0370-2693(83)90638-X} {\bibfield
  {journal} {\bibinfo  {journal} {Phys. Lett.}\ }\textbf {\bibinfo {volume}
  {B120}},\ \bibinfo {pages} {133} (\bibinfo {year} {1983})}\BibitemShut
  {NoStop}%
%%CITATION = PHLTA,B120,133;%%
\bibitem [{\citenamefont {Dine}\ and\ \citenamefont
  {Fischler}(1983)}]{Dine:1982ah}%
  \BibitemOpen
  \bibfield  {author} {\bibinfo {author} {\bibfnamefont {M.}~\bibnamefont
  {Dine}}\ and\ \bibinfo {author} {\bibfnamefont {W.}~\bibnamefont
  {Fischler}},\ }\href {\doibase 10.1016/0370-2693(83)90639-1} {\bibfield
  {journal} {\bibinfo  {journal} {Phys. Lett.}\ }\textbf {\bibinfo {volume}
  {B120}},\ \bibinfo {pages} {137} (\bibinfo {year} {1983})}\BibitemShut
  {NoStop}%
%%CITATION = PHLTA,B120,137;%%
\bibitem [{\citenamefont {Vilenkin}\ and\ \citenamefont
  {Everett}(1982)}]{Vilenkin:1982ks}%
  \BibitemOpen
  \bibfield  {author} {\bibinfo {author} {\bibfnamefont {A.}~\bibnamefont
  {Vilenkin}}\ and\ \bibinfo {author} {\bibfnamefont {A.}~\bibnamefont
  {Everett}},\ }\href {\doibase 10.1103/PhysRevLett.48.1867} {\bibfield
  {journal} {\bibinfo  {journal} {Phys. Rev. Lett.}\ }\textbf {\bibinfo
  {volume} {48}},\ \bibinfo {pages} {1867} (\bibinfo {year}
  {1982})}\BibitemShut {NoStop}%
\bibitem [{\citenamefont {Davis}(1986)}]{Davis:1986xc}%
  \BibitemOpen
  \bibfield  {author} {\bibinfo {author} {\bibfnamefont {R.~L.}\ \bibnamefont
  {Davis}},\ }\href {\doibase 10.1016/0370-2693(86)90300-X} {\bibfield
  {journal} {\bibinfo  {journal} {Phys. Lett.}\ }\textbf {\bibinfo {volume}
  {B180}},\ \bibinfo {pages} {225} (\bibinfo {year} {1986})}\BibitemShut
  {NoStop}%
%%CITATION = PHLTA,B180,225;%%
\bibitem [{\citenamefont {Hindmarsh}\ and\ \citenamefont
  {Kibble}(1995)}]{Hindmarsh:1994re}%
  \BibitemOpen
  \bibfield  {author} {\bibinfo {author} {\bibfnamefont {M.}~\bibnamefont
  {Hindmarsh}}\ and\ \bibinfo {author} {\bibfnamefont {T.}~\bibnamefont
  {Kibble}},\ }\href {\doibase 10.1088/0034-4885/58/5/001} {\bibfield
  {journal} {\bibinfo  {journal} {Rept.Prog.Phys.}\ }\textbf {\bibinfo {volume}
  {58}},\ \bibinfo {pages} {477} (\bibinfo {year} {1995})},\ \Eprint
  {http://arxiv.org/abs/hep-ph/9411342} {arXiv:hep-ph/9411342 [hep-ph]}
  \BibitemShut {NoStop}%
%%CITATION = HEP-PH/9411342;%%
\bibitem [{\citenamefont {Vilenkin}\ and\ \citenamefont
  {Shellard}(2000)}]{Vilenkin:2000jqa}%
  \BibitemOpen
  \bibfield  {author} {\bibinfo {author} {\bibfnamefont {A.}~\bibnamefont
  {Vilenkin}}\ and\ \bibinfo {author} {\bibfnamefont {E.~P.~S.}\ \bibnamefont
  {Shellard}},\ }\href
  {http://www.cambridge.org/mw/academic/subjects/physics/theoretical-physics-and-mathematical-physics/cosmic-strings-and-other-topological-defects?format=PB}
  {\emph {\bibinfo {title} {{Cosmic Strings and Other Topological Defects}}}}\
  (\bibinfo  {publisher} {Cambridge University Press},\ \bibinfo {year}
  {2000})\BibitemShut {NoStop}%
%%CITATION = INSPIRE-1384873;%%
\bibitem [{\citenamefont {Sikivie}(1982)}]{Sikivie:1982qv}%
  \BibitemOpen
  \bibfield  {author} {\bibinfo {author} {\bibfnamefont {P.}~\bibnamefont
  {Sikivie}},\ }\href {\doibase 10.1103/PhysRevLett.48.1156} {\bibfield
  {journal} {\bibinfo  {journal} {Phys. Rev. Lett.}\ }\textbf {\bibinfo
  {volume} {48}},\ \bibinfo {pages} {1156} (\bibinfo {year}
  {1982})}\BibitemShut {NoStop}%
%%CITATION = PRLTA,48,1156;%%
\bibitem [{\citenamefont {Georgi}\ and\ \citenamefont
  {Wise}(1982)}]{Georgi:1982ph}%
  \BibitemOpen
  \bibfield  {author} {\bibinfo {author} {\bibfnamefont {H.}~\bibnamefont
  {Georgi}}\ and\ \bibinfo {author} {\bibfnamefont {M.~B.}\ \bibnamefont
  {Wise}},\ }\href {\doibase 10.1016/0370-2693(82)90989-3} {\bibfield
  {journal} {\bibinfo  {journal} {Phys. Lett.}\ }\textbf {\bibinfo {volume}
  {116B}},\ \bibinfo {pages} {123} (\bibinfo {year} {1982})}\BibitemShut
  {NoStop}%
%%CITATION = PHLTA,116B,123;%%
\bibitem [{\citenamefont {Hogan}\ and\ \citenamefont
  {Rees}(1988)}]{Hogan:1988mp}%
  \BibitemOpen
  \bibfield  {author} {\bibinfo {author} {\bibfnamefont {C.}~\bibnamefont
  {Hogan}}\ and\ \bibinfo {author} {\bibfnamefont {M.}~\bibnamefont {Rees}},\
  }\href {\doibase 10.1016/0370-2693(88)91655-3} {\bibfield  {journal}
  {\bibinfo  {journal} {Phys. Lett. B}\ }\textbf {\bibinfo {volume} {205}},\
  \bibinfo {pages} {228} (\bibinfo {year} {1988})}\BibitemShut {NoStop}%
\bibitem [{\citenamefont {Kolb}\ and\ \citenamefont
  {Tkachev}(1993)}]{Kolb:1993zz}%
  \BibitemOpen
  \bibfield  {author} {\bibinfo {author} {\bibfnamefont {E.~W.}\ \bibnamefont
  {Kolb}}\ and\ \bibinfo {author} {\bibfnamefont {I.~I.}\ \bibnamefont
  {Tkachev}},\ }\href {\doibase 10.1103/PhysRevLett.71.3051} {\bibfield
  {journal} {\bibinfo  {journal} {Phys. Rev. Lett.}\ }\textbf {\bibinfo
  {volume} {71}},\ \bibinfo {pages} {3051} (\bibinfo {year} {1993})},\ \Eprint
  {http://arxiv.org/abs/hep-ph/9303313} {arXiv:hep-ph/9303313} \BibitemShut
  {NoStop}%
\bibitem [{\citenamefont {Kolb}\ and\ \citenamefont
  {Tkachev}(1994)}]{Kolb:1994fi}%
  \BibitemOpen
  \bibfield  {author} {\bibinfo {author} {\bibfnamefont {E.~W.}\ \bibnamefont
  {Kolb}}\ and\ \bibinfo {author} {\bibfnamefont {I.~I.}\ \bibnamefont
  {Tkachev}},\ }\href {\doibase 10.1103/PhysRevD.50.769} {\bibfield  {journal}
  {\bibinfo  {journal} {Phys. Rev. D}\ }\textbf {\bibinfo {volume} {50}},\
  \bibinfo {pages} {769} (\bibinfo {year} {1994})},\ \Eprint
  {http://arxiv.org/abs/astro-ph/9403011} {arXiv:astro-ph/9403011} \BibitemShut
  {NoStop}%
\bibitem [{\citenamefont {Kolb}\ and\ \citenamefont
  {Tkachev}(1996)}]{Kolb:1995bu}%
  \BibitemOpen
  \bibfield  {author} {\bibinfo {author} {\bibfnamefont {E.~W.}\ \bibnamefont
  {Kolb}}\ and\ \bibinfo {author} {\bibfnamefont {I.~I.}\ \bibnamefont
  {Tkachev}},\ }\href {\doibase 10.1086/309962} {\bibfield  {journal} {\bibinfo
   {journal} {Astrophys. J. Lett.}\ }\textbf {\bibinfo {volume} {460}},\
  \bibinfo {pages} {L25} (\bibinfo {year} {1996})},\ \Eprint
  {http://arxiv.org/abs/astro-ph/9510043} {arXiv:astro-ph/9510043} \BibitemShut
  {NoStop}%
\bibitem [{\citenamefont {Fairbairn}\ \emph {et~al.}(2017)\citenamefont
  {Fairbairn}, \citenamefont {Marsh},\ and\ \citenamefont
  {Quevillon}}]{Fairbairn:2017dmf}%
  \BibitemOpen
  \bibfield  {author} {\bibinfo {author} {\bibfnamefont {M.}~\bibnamefont
  {Fairbairn}}, \bibinfo {author} {\bibfnamefont {D.~J.~E.}\ \bibnamefont
  {Marsh}}, \ and\ \bibinfo {author} {\bibfnamefont {J.}~\bibnamefont
  {Quevillon}},\ }\href {\doibase 10.1103/PhysRevLett.119.021101} {\bibfield
  {journal} {\bibinfo  {journal} {Phys. Rev. Lett.}\ }\textbf {\bibinfo
  {volume} {119}},\ \bibinfo {pages} {021101} (\bibinfo {year} {2017})},\
  \Eprint {http://arxiv.org/abs/1701.04787} {arXiv:1701.04787 [astro-ph.CO]}
  \BibitemShut {NoStop}%
\bibitem [{\citenamefont {Fairbairn}\ \emph {et~al.}(2018)\citenamefont
  {Fairbairn}, \citenamefont {Marsh}, \citenamefont {Quevillon},\ and\
  \citenamefont {Rozier}}]{Fairbairn:2017sil}%
  \BibitemOpen
  \bibfield  {author} {\bibinfo {author} {\bibfnamefont {M.}~\bibnamefont
  {Fairbairn}}, \bibinfo {author} {\bibfnamefont {D.~J.~E.}\ \bibnamefont
  {Marsh}}, \bibinfo {author} {\bibfnamefont {J.}~\bibnamefont {Quevillon}}, \
  and\ \bibinfo {author} {\bibfnamefont {S.}~\bibnamefont {Rozier}},\ }\href
  {\doibase 10.1103/PhysRevD.97.083502} {\bibfield  {journal} {\bibinfo
  {journal} {Phys. Rev. D}\ }\textbf {\bibinfo {volume} {97}},\ \bibinfo
  {pages} {083502} (\bibinfo {year} {2018})},\ \Eprint
  {http://arxiv.org/abs/1707.03310} {arXiv:1707.03310 [astro-ph.CO]}
  \BibitemShut {NoStop}%
\bibitem [{\citenamefont {Yamaguchi}\ \emph {et~al.}(1999)\citenamefont
  {Yamaguchi}, \citenamefont {Kawasaki},\ and\ \citenamefont
  {Yokoyama}}]{Yamaguchi:1998gx}%
  \BibitemOpen
  \bibfield  {author} {\bibinfo {author} {\bibfnamefont {M.}~\bibnamefont
  {Yamaguchi}}, \bibinfo {author} {\bibfnamefont {M.}~\bibnamefont {Kawasaki}},
  \ and\ \bibinfo {author} {\bibfnamefont {J.}~\bibnamefont {Yokoyama}},\
  }\href {\doibase 10.1103/PhysRevLett.82.4578} {\bibfield  {journal} {\bibinfo
   {journal} {Phys. Rev. Lett.}\ }\textbf {\bibinfo {volume} {82}},\ \bibinfo
  {pages} {4578} (\bibinfo {year} {1999})},\ \Eprint
  {http://arxiv.org/abs/hep-ph/9811311} {arXiv:hep-ph/9811311 [hep-ph]}
  \BibitemShut {NoStop}%
%%CITATION = HEP-PH/9811311;%%
\bibitem [{\citenamefont {Yamaguchi}(1999)}]{Yamaguchi:1999yp}%
  \BibitemOpen
  \bibfield  {author} {\bibinfo {author} {\bibfnamefont {M.}~\bibnamefont
  {Yamaguchi}},\ }\href {\doibase 10.1103/PhysRevD.60.103511} {\bibfield
  {journal} {\bibinfo  {journal} {Phys. Rev. D}\ }\textbf {\bibinfo {volume}
  {60}},\ \bibinfo {pages} {103511} (\bibinfo {year} {1999})},\ \Eprint
  {http://arxiv.org/abs/hep-ph/9907506} {arXiv:hep-ph/9907506} \BibitemShut
  {NoStop}%
\bibitem [{\citenamefont {Yamaguchi}\ \emph {et~al.}(2000)\citenamefont
  {Yamaguchi}, \citenamefont {Yokoyama},\ and\ \citenamefont
  {Kawasaki}}]{Yamaguchi:1999dy}%
  \BibitemOpen
  \bibfield  {author} {\bibinfo {author} {\bibfnamefont {M.}~\bibnamefont
  {Yamaguchi}}, \bibinfo {author} {\bibfnamefont {J.}~\bibnamefont {Yokoyama}},
  \ and\ \bibinfo {author} {\bibfnamefont {M.}~\bibnamefont {Kawasaki}},\
  }\href {\doibase 10.1103/PhysRevD.61.061301} {\bibfield  {journal} {\bibinfo
  {journal} {Phys. Rev. D}\ }\textbf {\bibinfo {volume} {61}},\ \bibinfo
  {pages} {061301} (\bibinfo {year} {2000})},\ \Eprint
  {http://arxiv.org/abs/hep-ph/9910352} {arXiv:hep-ph/9910352} \BibitemShut
  {NoStop}%
\bibitem [{\citenamefont {Yamaguchi}\ and\ \citenamefont
  {Yokoyama}(2003)}]{Yamaguchi:2002sh}%
  \BibitemOpen
  \bibfield  {author} {\bibinfo {author} {\bibfnamefont {M.}~\bibnamefont
  {Yamaguchi}}\ and\ \bibinfo {author} {\bibfnamefont {J.}~\bibnamefont
  {Yokoyama}},\ }\href {\doibase 10.1103/PhysRevD.67.103514} {\bibfield
  {journal} {\bibinfo  {journal} {Phys. Rev.}\ }\textbf {\bibinfo {volume}
  {D67}},\ \bibinfo {pages} {103514} (\bibinfo {year} {2003})},\ \Eprint
  {http://arxiv.org/abs/hep-ph/0210343} {arXiv:hep-ph/0210343 [hep-ph]}
  \BibitemShut {NoStop}%
%%CITATION = HEP-PH/0210343;%%
\bibitem [{\citenamefont {Hiramatsu}\ \emph {et~al.}(2011)\citenamefont
  {Hiramatsu}, \citenamefont {Kawasaki}, \citenamefont {Sekiguchi},
  \citenamefont {Yamaguchi},\ and\ \citenamefont
  {Yokoyama}}]{Hiramatsu:2010yu}%
  \BibitemOpen
  \bibfield  {author} {\bibinfo {author} {\bibfnamefont {T.}~\bibnamefont
  {Hiramatsu}}, \bibinfo {author} {\bibfnamefont {M.}~\bibnamefont {Kawasaki}},
  \bibinfo {author} {\bibfnamefont {T.}~\bibnamefont {Sekiguchi}}, \bibinfo
  {author} {\bibfnamefont {M.}~\bibnamefont {Yamaguchi}}, \ and\ \bibinfo
  {author} {\bibfnamefont {J.}~\bibnamefont {Yokoyama}},\ }\href {\doibase
  10.1103/PhysRevD.83.123531} {\bibfield  {journal} {\bibinfo  {journal} {Phys.
  Rev.}\ }\textbf {\bibinfo {volume} {D83}},\ \bibinfo {pages} {123531}
  (\bibinfo {year} {2011})},\ \Eprint {http://arxiv.org/abs/1012.5502}
  {arXiv:1012.5502 [hep-ph]} \BibitemShut {NoStop}%
%%CITATION = ARXIV:1012.5502;%%
\bibitem [{\citenamefont {Hiramatsu}\ \emph {et~al.}(2012)\citenamefont
  {Hiramatsu}, \citenamefont {Kawasaki}, \citenamefont {Saikawa},\ and\
  \citenamefont {Sekiguchi}}]{Hiramatsu:2012gg}%
  \BibitemOpen
  \bibfield  {author} {\bibinfo {author} {\bibfnamefont {T.}~\bibnamefont
  {Hiramatsu}}, \bibinfo {author} {\bibfnamefont {M.}~\bibnamefont {Kawasaki}},
  \bibinfo {author} {\bibfnamefont {K.}~\bibnamefont {Saikawa}}, \ and\
  \bibinfo {author} {\bibfnamefont {T.}~\bibnamefont {Sekiguchi}},\ }\href
  {\doibase 10.1103/PhysRevD.86.089902, 10.1103/PhysRevD.85.105020} {\bibfield
  {journal} {\bibinfo  {journal} {Phys. Rev.}\ }\textbf {\bibinfo {volume}
  {D85}},\ \bibinfo {pages} {105020} (\bibinfo {year} {2012})},\ \bibinfo
  {note} {[Erratum: Phys. Rev.D86,089902(2012)]},\ \Eprint
  {http://arxiv.org/abs/1202.5851} {arXiv:1202.5851 [hep-ph]} \BibitemShut
  {NoStop}%
%%CITATION = ARXIV:1202.5851;%%
\bibitem [{\citenamefont {Kawasaki}\ \emph {et~al.}(2015)\citenamefont
  {Kawasaki}, \citenamefont {Saikawa},\ and\ \citenamefont
  {Sekiguchi}}]{Kawasaki:2014sqa}%
  \BibitemOpen
  \bibfield  {author} {\bibinfo {author} {\bibfnamefont {M.}~\bibnamefont
  {Kawasaki}}, \bibinfo {author} {\bibfnamefont {K.}~\bibnamefont {Saikawa}}, \
  and\ \bibinfo {author} {\bibfnamefont {T.}~\bibnamefont {Sekiguchi}},\ }\href
  {\doibase 10.1103/PhysRevD.91.065014} {\bibfield  {journal} {\bibinfo
  {journal} {Phys. Rev.}\ }\textbf {\bibinfo {volume} {D91}},\ \bibinfo {pages}
  {065014} (\bibinfo {year} {2015})},\ \Eprint {http://arxiv.org/abs/1412.0789}
  {arXiv:1412.0789 [hep-ph]} \BibitemShut {NoStop}%
%%CITATION = ARXIV:1412.0789;%%
\bibitem [{\citenamefont {Fleury}\ and\ \citenamefont
  {Moore}(2016)}]{Fleury:2015aca}%
  \BibitemOpen
  \bibfield  {author} {\bibinfo {author} {\bibfnamefont {L.}~\bibnamefont
  {Fleury}}\ and\ \bibinfo {author} {\bibfnamefont {G.~D.}\ \bibnamefont
  {Moore}},\ }\href {\doibase 10.1088/1475-7516/2016/01/004} {\bibfield
  {journal} {\bibinfo  {journal} {JCAP}\ }\textbf {\bibinfo {volume} {1601}},\
  \bibinfo {pages} {004} (\bibinfo {year} {2016})},\ \Eprint
  {http://arxiv.org/abs/1509.00026} {arXiv:1509.00026 [hep-ph]} \BibitemShut
  {NoStop}%
%%CITATION = ARXIV:1509.00026;%%
\bibitem [{\citenamefont {Lopez-Eiguren}\ \emph {et~al.}(2017)\citenamefont
  {Lopez-Eiguren}, \citenamefont {Lizarraga}, \citenamefont {Hindmarsh},\ and\
  \citenamefont {Urrestilla}}]{Lopez-Eiguren:2017dmc}%
  \BibitemOpen
  \bibfield  {author} {\bibinfo {author} {\bibfnamefont {A.}~\bibnamefont
  {Lopez-Eiguren}}, \bibinfo {author} {\bibfnamefont {J.}~\bibnamefont
  {Lizarraga}}, \bibinfo {author} {\bibfnamefont {M.}~\bibnamefont
  {Hindmarsh}}, \ and\ \bibinfo {author} {\bibfnamefont {J.}~\bibnamefont
  {Urrestilla}},\ }\href {\doibase 10.1088/1475-7516/2017/07/026} {\bibfield
  {journal} {\bibinfo  {journal} {JCAP}\ }\textbf {\bibinfo {volume} {1707}},\
  \bibinfo {pages} {026} (\bibinfo {year} {2017})},\ \Eprint
  {http://arxiv.org/abs/1705.04154} {arXiv:1705.04154 [astro-ph.CO]}
  \BibitemShut {NoStop}%
%%CITATION = ARXIV:1705.04154;%%
\bibitem [{\citenamefont {Klaer}\ and\ \citenamefont
  {Moore}(2017{\natexlab{a}})}]{Klaer:2017qhr}%
  \BibitemOpen
  \bibfield  {author} {\bibinfo {author} {\bibfnamefont {V.~B.}\ \bibnamefont
  {Klaer}}\ and\ \bibinfo {author} {\bibfnamefont {G.~D.}\ \bibnamefont
  {Moore}},\ }\href {\doibase 10.1088/1475-7516/2017/10/043} {\bibfield
  {journal} {\bibinfo  {journal} {JCAP}\ }\textbf {\bibinfo {volume} {1710}},\
  \bibinfo {pages} {043} (\bibinfo {year} {2017}{\natexlab{a}})},\ \Eprint
  {http://arxiv.org/abs/1707.05566} {arXiv:1707.05566 [hep-ph]} \BibitemShut
  {NoStop}%
%%CITATION = ARXIV:1707.05566;%%
\bibitem [{\citenamefont {Klaer}\ and\ \citenamefont
  {Moore}(2017{\natexlab{b}})}]{Klaer:2017ond}%
  \BibitemOpen
  \bibfield  {author} {\bibinfo {author} {\bibfnamefont {V.~B.}\ \bibnamefont
  {Klaer}}\ and\ \bibinfo {author} {\bibfnamefont {G.~D.}\ \bibnamefont
  {Moore}},\ }\href {\doibase 10.1088/1475-7516/2017/11/049} {\bibfield
  {journal} {\bibinfo  {journal} {JCAP}\ }\textbf {\bibinfo {volume} {1711}},\
  \bibinfo {pages} {049} (\bibinfo {year} {2017}{\natexlab{b}})},\ \Eprint
  {http://arxiv.org/abs/1708.07521} {arXiv:1708.07521 [hep-ph]} \BibitemShut
  {NoStop}%
%%CITATION = ARXIV:1708.07521;%%
\bibitem [{\citenamefont {Gorghetto}\ \emph {et~al.}(2018)\citenamefont
  {Gorghetto}, \citenamefont {Hardy},\ and\ \citenamefont
  {Villadoro}}]{Gorghetto:2018myk}%
  \BibitemOpen
  \bibfield  {author} {\bibinfo {author} {\bibfnamefont {M.}~\bibnamefont
  {Gorghetto}}, \bibinfo {author} {\bibfnamefont {E.}~\bibnamefont {Hardy}}, \
  and\ \bibinfo {author} {\bibfnamefont {G.}~\bibnamefont {Villadoro}},\ }\href
  {\doibase 10.1007/JHEP07(2018)151} {\bibfield  {journal} {\bibinfo  {journal}
  {JHEP}\ }\textbf {\bibinfo {volume} {07}},\ \bibinfo {pages} {151} (\bibinfo
  {year} {2018})},\ \Eprint {http://arxiv.org/abs/1806.04677} {arXiv:1806.04677
  [hep-ph]} \BibitemShut {NoStop}%
%%CITATION = ARXIV:1806.04677;%%
\bibitem [{\citenamefont {Kawasaki}\ \emph {et~al.}(2018)\citenamefont
  {Kawasaki}, \citenamefont {Sekiguchi}, \citenamefont {Yamaguchi},\ and\
  \citenamefont {Yokoyama}}]{Kawasaki:2018bzv}%
  \BibitemOpen
  \bibfield  {author} {\bibinfo {author} {\bibfnamefont {M.}~\bibnamefont
  {Kawasaki}}, \bibinfo {author} {\bibfnamefont {T.}~\bibnamefont {Sekiguchi}},
  \bibinfo {author} {\bibfnamefont {M.}~\bibnamefont {Yamaguchi}}, \ and\
  \bibinfo {author} {\bibfnamefont {J.}~\bibnamefont {Yokoyama}},\ }\href
  {\doibase 10.1093/ptep/pty098} {\bibfield  {journal} {\bibinfo  {journal}
  {PTEP}\ }\textbf {\bibinfo {volume} {2018}},\ \bibinfo {pages} {091E01}
  (\bibinfo {year} {2018})},\ \Eprint {http://arxiv.org/abs/1806.05566}
  {arXiv:1806.05566 [hep-ph]} \BibitemShut {NoStop}%
%%CITATION = ARXIV:1806.05566;%%
\bibitem [{\citenamefont {Vaquero}\ \emph {et~al.}(2019)\citenamefont
  {Vaquero}, \citenamefont {Redondo},\ and\ \citenamefont
  {Stadler}}]{Vaquero:2018tib}%
  \BibitemOpen
  \bibfield  {author} {\bibinfo {author} {\bibfnamefont {A.}~\bibnamefont
  {Vaquero}}, \bibinfo {author} {\bibfnamefont {J.}~\bibnamefont {Redondo}}, \
  and\ \bibinfo {author} {\bibfnamefont {J.}~\bibnamefont {Stadler}},\ }\href
  {\doibase 10.1088/1475-7516/2019/04/012} {\bibfield  {journal} {\bibinfo
  {journal} {JCAP}\ }\textbf {\bibinfo {volume} {1904}},\ \bibinfo {pages}
  {012} (\bibinfo {year} {2019})},\ \Eprint {http://arxiv.org/abs/1809.09241}
  {arXiv:1809.09241 [astro-ph.CO]} \BibitemShut {NoStop}%
\bibitem [{\citenamefont {Buschmann}\ \emph {et~al.}(2020)\citenamefont
  {Buschmann}, \citenamefont {Foster},\ and\ \citenamefont
  {Safdi}}]{Buschmann:2019icd}%
  \BibitemOpen
  \bibfield  {author} {\bibinfo {author} {\bibfnamefont {M.}~\bibnamefont
  {Buschmann}}, \bibinfo {author} {\bibfnamefont {J.~W.}\ \bibnamefont
  {Foster}}, \ and\ \bibinfo {author} {\bibfnamefont {B.~R.}\ \bibnamefont
  {Safdi}},\ }\href {\doibase 10.1103/PhysRevLett.124.161103} {\bibfield
  {journal} {\bibinfo  {journal} {Phys. Rev. Lett.}\ }\textbf {\bibinfo
  {volume} {124}},\ \bibinfo {pages} {161103} (\bibinfo {year} {2020})},\
  \Eprint {http://arxiv.org/abs/1906.00967} {arXiv:1906.00967 [astro-ph.CO]}
  \BibitemShut {NoStop}%
\bibitem [{\citenamefont {Klaer}\ and\ \citenamefont
  {Moore}(2020)}]{Klaer:2019fxc}%
  \BibitemOpen
  \bibfield  {author} {\bibinfo {author} {\bibfnamefont {V.~B.}\ \bibnamefont
  {Klaer}}\ and\ \bibinfo {author} {\bibfnamefont {G.~D.}\ \bibnamefont
  {Moore}},\ }\href {\doibase 10.1088/1475-7516/2020/06/021} {\bibfield
  {journal} {\bibinfo  {journal} {JCAP}\ }\textbf {\bibinfo {volume} {06}},\
  \bibinfo {pages} {021} (\bibinfo {year} {2020})},\ \Eprint
  {http://arxiv.org/abs/1912.08058} {arXiv:1912.08058 [hep-ph]} \BibitemShut
  {NoStop}%
\bibitem [{\citenamefont {Hindmarsh}\ \emph {et~al.}(2020)\citenamefont
  {Hindmarsh}, \citenamefont {Lizarraga}, \citenamefont {Lopez-Eiguren},\ and\
  \citenamefont {Urrestilla}}]{Hindmarsh:2019csc}%
  \BibitemOpen
  \bibfield  {author} {\bibinfo {author} {\bibfnamefont {M.}~\bibnamefont
  {Hindmarsh}}, \bibinfo {author} {\bibfnamefont {J.}~\bibnamefont
  {Lizarraga}}, \bibinfo {author} {\bibfnamefont {A.}~\bibnamefont
  {Lopez-Eiguren}}, \ and\ \bibinfo {author} {\bibfnamefont {J.}~\bibnamefont
  {Urrestilla}},\ }\href {\doibase 10.1103/PhysRevLett.124.021301} {\bibfield
  {journal} {\bibinfo  {journal} {Phys. Rev. Lett.}\ }\textbf {\bibinfo
  {volume} {124}},\ \bibinfo {pages} {021301} (\bibinfo {year} {2020})},\
  \Eprint {http://arxiv.org/abs/1908.03522} {arXiv:1908.03522 [astro-ph.CO]}
  \BibitemShut {NoStop}%
\bibitem [{\citenamefont {Gorghetto}\ \emph {et~al.}(2020)\citenamefont
  {Gorghetto}, \citenamefont {Hardy},\ and\ \citenamefont
  {Villadoro}}]{Gorghetto:2020qws}%
  \BibitemOpen
  \bibfield  {author} {\bibinfo {author} {\bibfnamefont {M.}~\bibnamefont
  {Gorghetto}}, \bibinfo {author} {\bibfnamefont {E.}~\bibnamefont {Hardy}}, \
  and\ \bibinfo {author} {\bibfnamefont {G.}~\bibnamefont {Villadoro}},\
  }\href@noop {} {\  (\bibinfo {year} {2020})},\ \Eprint
  {http://arxiv.org/abs/2007.04990} {arXiv:2007.04990 [hep-ph]} \BibitemShut
  {NoStop}%
\bibitem [{\citenamefont {Gorghetto}\ \emph {et~al.}(2021)\citenamefont
  {Gorghetto}, \citenamefont {Hardy},\ and\ \citenamefont
  {Nicolaescu}}]{Gorghetto:2021fsn}%
  \BibitemOpen
  \bibfield  {author} {\bibinfo {author} {\bibfnamefont {M.}~\bibnamefont
  {Gorghetto}}, \bibinfo {author} {\bibfnamefont {E.}~\bibnamefont {Hardy}}, \
  and\ \bibinfo {author} {\bibfnamefont {H.}~\bibnamefont {Nicolaescu}},\
  }\href@noop {} {\  (\bibinfo {year} {2021})},\ \Eprint
  {http://arxiv.org/abs/2101.11007} {arXiv:2101.11007 [hep-ph]} \BibitemShut
  {NoStop}%
\bibitem [{\citenamefont {Dine}\ \emph {et~al.}(2020)\citenamefont {Dine},
  \citenamefont {Fernandez}, \citenamefont {Ghalsasi},\ and\ \citenamefont
  {Patel}}]{Dine:2020pds}%
  \BibitemOpen
  \bibfield  {author} {\bibinfo {author} {\bibfnamefont {M.}~\bibnamefont
  {Dine}}, \bibinfo {author} {\bibfnamefont {N.}~\bibnamefont {Fernandez}},
  \bibinfo {author} {\bibfnamefont {A.}~\bibnamefont {Ghalsasi}}, \ and\
  \bibinfo {author} {\bibfnamefont {H.~H.}\ \bibnamefont {Patel}},\ }\href@noop
  {} {\  (\bibinfo {year} {2020})},\ \Eprint {http://arxiv.org/abs/2012.13065}
  {arXiv:2012.13065 [hep-ph]} \BibitemShut {NoStop}%
\bibitem [{\citenamefont {Braine}\ \emph {et~al.}(2020)\citenamefont {Braine}
  \emph {et~al.}}]{Braine:2019fqb}%
  \BibitemOpen
  \bibfield  {author} {\bibinfo {author} {\bibfnamefont {T.}~\bibnamefont
  {Braine}} \emph {et~al.} (\bibinfo {collaboration} {ADMX}),\ }\href {\doibase
  10.1103/PhysRevLett.124.101303} {\bibfield  {journal} {\bibinfo  {journal}
  {Phys. Rev. Lett.}\ }\textbf {\bibinfo {volume} {124}},\ \bibinfo {pages}
  {101303} (\bibinfo {year} {2020})},\ \Eprint
  {http://arxiv.org/abs/1910.08638} {arXiv:1910.08638 [hep-ex]} \BibitemShut
  {NoStop}%
\bibitem [{\citenamefont {Kibble}(1985)}]{Kibble:1984hp}%
  \BibitemOpen
  \bibfield  {author} {\bibinfo {author} {\bibfnamefont {T.~W.~B.}\
  \bibnamefont {Kibble}},\ }\bibfield  {booktitle} {\emph {\bibinfo {booktitle}
  {{Phase transitions in the very early Univesre. proceedings, International
  Workshop, Bielefeld, F.R. Germany, June 4-8, 1984}}},\ }\href {\doibase
  10.1016/0550-3213(85)90439-0, 10.1016/0550-3213(85)90596-6} {\bibfield
  {journal} {\bibinfo  {journal} {Nucl. Phys.}\ }\textbf {\bibinfo {volume}
  {B252}},\ \bibinfo {pages} {227} (\bibinfo {year} {1985})},\ \bibinfo {note}
  {[Erratum: Nucl. Phys.B261,750(1985)]}\BibitemShut {NoStop}%
%%CITATION = NUPHA,B252,227;%%
\bibitem [{\citenamefont {Martins}\ and\ \citenamefont
  {Shellard}(1996)}]{Martins:1996jp}%
  \BibitemOpen
  \bibfield  {author} {\bibinfo {author} {\bibfnamefont {C.~J. A.~P.}\
  \bibnamefont {Martins}}\ and\ \bibinfo {author} {\bibfnamefont {E.~P.~S.}\
  \bibnamefont {Shellard}},\ }\href {\doibase 10.1103/PhysRevD.54.2535}
  {\bibfield  {journal} {\bibinfo  {journal} {Phys. Rev.}\ }\textbf {\bibinfo
  {volume} {D54}},\ \bibinfo {pages} {2535} (\bibinfo {year} {1996})},\ \Eprint
  {http://arxiv.org/abs/hep-ph/9602271} {arXiv:hep-ph/9602271 [hep-ph]}
  \BibitemShut {NoStop}%
%%CITATION = HEP-PH/9602271;%%
\bibitem [{\citenamefont {Martins}\ and\ \citenamefont
  {Shellard}(2002)}]{Martins:2000cs}%
  \BibitemOpen
  \bibfield  {author} {\bibinfo {author} {\bibfnamefont {C.}~\bibnamefont
  {Martins}}\ and\ \bibinfo {author} {\bibfnamefont {E.}~\bibnamefont
  {Shellard}},\ }\href {\doibase 10.1103/PhysRevD.65.043514} {\bibfield
  {journal} {\bibinfo  {journal} {Phys.Rev.}\ }\textbf {\bibinfo {volume}
  {D65}},\ \bibinfo {pages} {043514} (\bibinfo {year} {2002})},\ \Eprint
  {http://arxiv.org/abs/hep-ph/0003298} {arXiv:hep-ph/0003298 [hep-ph]}
  \BibitemShut {NoStop}%
%%CITATION = HEP-PH/0003298;%%
\bibitem [{\citenamefont {Martins}(2019)}]{Martins:2018dqg}%
  \BibitemOpen
  \bibfield  {author} {\bibinfo {author} {\bibfnamefont {C.~J. A.~P.}\
  \bibnamefont {Martins}},\ }\href {\doibase 10.1016/j.physletb.2018.11.031}
  {\bibfield  {journal} {\bibinfo  {journal} {Phys. Lett.}\ }\textbf {\bibinfo
  {volume} {B788}},\ \bibinfo {pages} {147} (\bibinfo {year} {2019})},\ \Eprint
  {http://arxiv.org/abs/1811.12678} {arXiv:1811.12678 [astro-ph.CO]}
  \BibitemShut {NoStop}%
%%CITATION = ARXIV:1811.12678;%%
\bibitem [{\citenamefont {Kibble}(1976)}]{Kibble:1976sj}%
  \BibitemOpen
  \bibfield  {author} {\bibinfo {author} {\bibfnamefont {T.}~\bibnamefont
  {Kibble}},\ }\href {\doibase 10.1088/0305-4470/9/8/029} {\bibfield  {journal}
  {\bibinfo  {journal} {J.Phys.}\ }\textbf {\bibinfo {volume} {A9}},\ \bibinfo
  {pages} {1387} (\bibinfo {year} {1976})}\BibitemShut {NoStop}%
%%CITATION = ICTP/75/5 ETC.;%%
\bibitem [{\citenamefont {Zurek}(1996)}]{Zurek:1996sj}%
  \BibitemOpen
  \bibfield  {author} {\bibinfo {author} {\bibfnamefont {W.}~\bibnamefont
  {Zurek}},\ }\href {\doibase 10.1016/S0370-1573(96)00009-9} {\bibfield
  {journal} {\bibinfo  {journal} {Phys. Rept.}\ }\textbf {\bibinfo {volume}
  {276}},\ \bibinfo {pages} {177} (\bibinfo {year} {1996})},\ \Eprint
  {http://arxiv.org/abs/cond-mat/9607135} {arXiv:cond-mat/9607135} \BibitemShut
  {NoStop}%
\bibitem [{\citenamefont {Press}\ \emph {et~al.}(1989)\citenamefont {Press},
  \citenamefont {Ryden},\ and\ \citenamefont {Spergel}}]{Press:1989yh}%
  \BibitemOpen
  \bibfield  {author} {\bibinfo {author} {\bibfnamefont {W.~H.}\ \bibnamefont
  {Press}}, \bibinfo {author} {\bibfnamefont {B.~S.}\ \bibnamefont {Ryden}}, \
  and\ \bibinfo {author} {\bibfnamefont {D.~N.}\ \bibnamefont {Spergel}},\
  }\href {\doibase 10.1086/168151} {\bibfield  {journal} {\bibinfo  {journal}
  {Astrophys. J.}\ }\textbf {\bibinfo {volume} {347}},\ \bibinfo {pages} {590}
  (\bibinfo {year} {1989})}\BibitemShut {NoStop}%
%%CITATION = ASJOA,347,590;%%
\bibitem [{\citenamefont {Vachaspati}\ and\ \citenamefont
  {Vilenkin}(1984)}]{Vachaspati:1984dz}%
  \BibitemOpen
  \bibfield  {author} {\bibinfo {author} {\bibfnamefont {T.}~\bibnamefont
  {Vachaspati}}\ and\ \bibinfo {author} {\bibfnamefont {A.}~\bibnamefont
  {Vilenkin}},\ }\href {\doibase 10.1103/PhysRevD.30.2036} {\bibfield
  {journal} {\bibinfo  {journal} {Phys. Rev.}\ }\textbf {\bibinfo {volume}
  {D30}},\ \bibinfo {pages} {2036} (\bibinfo {year} {1984})}\BibitemShut
  {NoStop}%
%%CITATION = PHRVA,D30,2036;%%
\bibitem [{\citenamefont {Hindmarsh}\ \emph {et~al.}(2017)\citenamefont
  {Hindmarsh}, \citenamefont {Lizarraga}, \citenamefont {Urrestilla},
  \citenamefont {Daverio},\ and\ \citenamefont {Kunz}}]{Hindmarsh:2017qff}%
  \BibitemOpen
  \bibfield  {author} {\bibinfo {author} {\bibfnamefont {M.}~\bibnamefont
  {Hindmarsh}}, \bibinfo {author} {\bibfnamefont {J.}~\bibnamefont
  {Lizarraga}}, \bibinfo {author} {\bibfnamefont {J.}~\bibnamefont
  {Urrestilla}}, \bibinfo {author} {\bibfnamefont {D.}~\bibnamefont {Daverio}},
  \ and\ \bibinfo {author} {\bibfnamefont {M.}~\bibnamefont {Kunz}},\ }\href
  {\doibase 10.1103/PhysRevD.96.023525} {\bibfield  {journal} {\bibinfo
  {journal} {Phys. Rev.}\ }\textbf {\bibinfo {volume} {D96}},\ \bibinfo {pages}
  {023525} (\bibinfo {year} {2017})},\ \Eprint
  {http://arxiv.org/abs/1703.06696} {arXiv:1703.06696 [astro-ph.CO]}
  \BibitemShut {NoStop}%
%%CITATION = ARXIV:1703.06696;%%
\bibitem [{\citenamefont {Correia}\ and\ \citenamefont
  {Martins}(2019)}]{Correia:2019bdl}%
  \BibitemOpen
  \bibfield  {author} {\bibinfo {author} {\bibfnamefont {J.~R. C. C.~C.}\
  \bibnamefont {Correia}}\ and\ \bibinfo {author} {\bibfnamefont {C.~J. A.~P.}\
  \bibnamefont {Martins}},\ }\href {\doibase 10.1103/PhysRevD.100.103517}
  {\bibfield  {journal} {\bibinfo  {journal} {Phys. Rev. D}\ }\textbf {\bibinfo
  {volume} {100}},\ \bibinfo {pages} {103517} (\bibinfo {year} {2019})},\
  \Eprint {http://arxiv.org/abs/1911.03163} {arXiv:1911.03163 [astro-ph.CO]}
  \BibitemShut {NoStop}%
\bibitem [{\citenamefont {Svrcek}\ and\ \citenamefont
  {Witten}(2006)}]{Svrcek:2006yi}%
  \BibitemOpen
  \bibfield  {author} {\bibinfo {author} {\bibfnamefont {P.}~\bibnamefont
  {Svrcek}}\ and\ \bibinfo {author} {\bibfnamefont {E.}~\bibnamefont
  {Witten}},\ }\href {\doibase 10.1088/1126-6708/2006/06/051} {\bibfield
  {journal} {\bibinfo  {journal} {JHEP}\ }\textbf {\bibinfo {volume} {06}},\
  \bibinfo {pages} {051} (\bibinfo {year} {2006})},\ \Eprint
  {http://arxiv.org/abs/hep-th/0605206} {arXiv:hep-th/0605206} \BibitemShut
  {NoStop}%
\bibitem [{\citenamefont {Figueroa}\ \emph {et~al.}(2013)\citenamefont
  {Figueroa}, \citenamefont {Hindmarsh},\ and\ \citenamefont
  {Urrestilla}}]{Figueroa:2012kw}%
  \BibitemOpen
  \bibfield  {author} {\bibinfo {author} {\bibfnamefont {D.~G.}\ \bibnamefont
  {Figueroa}}, \bibinfo {author} {\bibfnamefont {M.}~\bibnamefont {Hindmarsh}},
  \ and\ \bibinfo {author} {\bibfnamefont {J.}~\bibnamefont {Urrestilla}},\
  }\href {\doibase 10.1103/PhysRevLett.110.101302} {\bibfield  {journal}
  {\bibinfo  {journal} {Phys. Rev. Lett.}\ }\textbf {\bibinfo {volume} {110}},\
  \bibinfo {pages} {101302} (\bibinfo {year} {2013})},\ \Eprint
  {http://arxiv.org/abs/1212.5458} {arXiv:1212.5458 [astro-ph.CO]} \BibitemShut
  {NoStop}%
\bibitem [{\citenamefont {Figueroa}\ \emph {et~al.}(2020)\citenamefont
  {Figueroa}, \citenamefont {Hindmarsh}, \citenamefont {Lizarraga},\ and\
  \citenamefont {Urrestilla}}]{Figueroa:2020lvo}%
  \BibitemOpen
  \bibfield  {author} {\bibinfo {author} {\bibfnamefont {D.~G.}\ \bibnamefont
  {Figueroa}}, \bibinfo {author} {\bibfnamefont {M.}~\bibnamefont {Hindmarsh}},
  \bibinfo {author} {\bibfnamefont {J.}~\bibnamefont {Lizarraga}}, \ and\
  \bibinfo {author} {\bibfnamefont {J.}~\bibnamefont {Urrestilla}},\ }\href
  {\doibase 10.1103/PhysRevD.102.103516} {\bibfield  {journal} {\bibinfo
  {journal} {Phys. Rev. D}\ }\textbf {\bibinfo {volume} {102}},\ \bibinfo
  {pages} {103516} (\bibinfo {year} {2020})},\ \Eprint
  {http://arxiv.org/abs/2007.03337} {arXiv:2007.03337 [astro-ph.CO]}
  \BibitemShut {NoStop}%
\end{thebibliography}%

\end{document}